\documentclass{article}
\pdfoutput=1

\PassOptionsToPackage{numbers,compress}{natbib}
\PassOptionsToPackage{table}{xcolor}
\PassOptionsToPackage{disable}{microtype}

\usepackage[preprint]{neurips_2026}

\usepackage{charter}                                                   %

\usepackage{xcolor}
\definecolor{rnAccent}{HTML}{2E7D6F}      %
\definecolor{rnAccentDark}{HTML}{246858}  %
\definecolor{rnAccentLight}{HTML}{E8F5F0} %
\definecolor{rnDark}{HTML}{1F2328}        %
\definecolor{rnSlate}{HTML}{6B7280}

\usepackage{hyperref}
\hypersetup{
  colorlinks=true,
  linkcolor=rnAccentDark,
  citecolor=rnAccentDark,
  urlcolor=rnAccentDark,
  pdfborder={0 0 0}
}
\usepackage{tcolorbox}
\tcbuselibrary{skins,breakable}
\usepackage{graphicx}
\usepackage{booktabs}

\makeatletter
\newlength{\rnHeroSpacing}
\newcommand{\rn@gap}{\par\vspace{\rnHeroSpacing}}
\newcommand{\rn@smallgap}{\par\vspace{0.67\rnHeroSpacing}}

\def\rn@code{}
\def\rn@website{}
\def\rn@heroabstract{}
\def\rn@affiliations{}
\newcommand{\code}[1]{\def\rn@code{#1}}
\newcommand{\website}[1]{\def\rn@website{#1}}
\newcommand{\heroabstract}[1]{\def\rn@heroabstract{#1}}
\newcommand{\affiliations}[1]{\def\rn@affiliations{#1}}

\newcommand*{\rn@authorsep}{\hspace{0.45em}\textperiodcentered\hspace{0.45em}}

\renewcommand{\@maketitle}{%
  \vbox{%
    \hsize\textwidth
    \linewidth\hsize
    \vskip 0.02in
    \begin{tcolorbox}[
      enhanced,
      colback=rnAccent, opacityback=0.10,
      colframe=rnAccent, opacityframe=0.10,
      boxrule=0pt, arc=6pt,
      left=20pt, right=20pt, top=12pt, bottom=10pt,
      before skip=0pt, after skip=0pt,
      width=\linewidth,
    ]
      \ifx\@title\@empty\else
        \begin{center}
          {\LARGE\bfseries\color{rnDark}%
           \hyphenpenalty=10000\exhyphenpenalty=10000%
           \@title\par}%
        \end{center}
        \rn@smallgap
      \fi
      \ifx\@author\@empty\else
        \begin{center}
          \begingroup
            \def\and{\rn@authorsep}%
            \def\And{\rn@authorsep}%
            \def\AND{\rn@authorsep}%
            {\color{rnDark}\@author\par}%
          \endgroup
        \end{center}
        \rn@smallgap
      \fi
      \ifx\rn@affiliations\@empty\else
        \begin{center}
          {\small\color{rnDark}\rn@affiliations\par}%
        \end{center}
        \rn@smallgap
      \fi
      {\color{rnDark!25}\rule{\linewidth}{0.4pt}}
      \rn@gap
      \ifx\rn@heroabstract\@empty\else
        {\small\color{rnDark}\rn@heroabstract\par}%
        \rn@gap
      \fi
      {\color{rnDark!25}\rule{\linewidth}{0.4pt}}
      \rn@gap
      \def\rn@hasmeta{0}%
      \ifx\rn@code\@empty\else\def\rn@hasmeta{1}\fi
      \ifx\rn@website\@empty\else\def\rn@hasmeta{1}\fi
      \ifx\@date\@empty\else\def\rn@hasmeta{1}\fi
      \if\rn@hasmeta1
        \begingroup
          \setlength{\parskip}{0pt}%
          \setlength{\baselineskip}{14pt}%
          \raggedright\small\color{rnDark}%
          \def\rn@first{1}%
          \ifx\rn@code\@empty\else
            \noindent\textbf{Code:}~\href{\rn@code}{\rn@code}%
            \def\rn@first{0}%
          \fi
          \ifx\rn@website\@empty\else
            \if\rn@first0\\\fi
            \textbf{Website:}~\href{\rn@website}{\rn@website}%
            \def\rn@first{0}%
          \fi
          \ifx\@date\@empty\else
            \if\rn@first0\\\fi
            \textbf{Date:}~\@date%
          \fi
          \par
        \endgroup
      \fi
    \end{tcolorbox}
    \vskip 0.12in \@minus 0.05in
  }
}

\newtcolorbox{rnnote}[1][]{
  enhanced, breakable,
  colback=rnAccentLight, colframe=rnAccentLight,
  fonttitle=\sffamily\bfseries,
  coltitle=rnDark,
  boxrule=0pt, arc=2pt,
  left=10pt, right=10pt, top=8pt, bottom=8pt,
  before skip=10pt, after skip=10pt,
  #1
}
\makeatother

\usepackage{subfigure}
\usepackage[utf8]{inputenc}
\usepackage[T1]{fontenc}
\usepackage{url}
\usepackage{nicefrac}
\usepackage{listings}
\usepackage{amsmath}
\usepackage{amssymb}
\usepackage[capitalize,noabbrev]{cleveref}
\usepackage{xspace}
\usepackage{makecell}
\usepackage{textcomp}
\usepackage{siunitx}
\usepackage{wrapfig}
\usepackage{placeins}  %
\usepackage[subtle]{savetrees}
\usepackage[textsize=tiny]{todonotes}

\usepackage{amsmath,amsfonts,bm}

\def\eqref#1{equation~\ref{#1}}

\def\1{\bm{1}}

\DeclareMathAlphabet{\mathsfit}{\encodingdefault}{\sfdefault}{m}{sl}
\SetMathAlphabet{\mathsfit}{bold}{\encodingdefault}{\sfdefault}{bx}{n}

\definecolor{darkgreen}{RGB}{0,100,0}
\definecolor{darkred}{RGB}{139,0,0}
\definecolor{clrgp}{rgb}{.9,0,.9}
\definecolor{red}{rgb}{.8,0,0}
\definecolor{blue}{rgb}{0,0, 0.8}
\definecolor{gray}{rgb}{0.41, 0.41, 0.41}
\definecolor{forestgreen}{rgb}{0.13, 0.55, 0.13}
\definecolor{subtle}{RGB}{152,78,163}

\newcommand{\comments}[1]{}

\newcommand{\wildchat}{\textsc{Wildchat}\xspace}
\newcommand{\nr}{\textsc{NaturalReasoning}\xspace}

\newcommand{\supergpqa}{\textsc{SuperGPQA}\xspace}
\newcommand{\mmlupro}{\textsc{MMLU Pro}\xspace}
\newcommand{\gpqa}{\textsc{gpqa}\xspace}

\newcommand{\qwenfour}{\textsc{Qwen3-4B}\xspace}
\newcommand{\qweneight}{\textsc{Qwen3-8B}\xspace}
\newcommand{\qwenfourteen}{\textsc{Qwen3-14B}\xspace}
\newcommand{\qwenthirtytwo}{\textsc{Qwen3-32B}\xspace}
\newcommand{\qwentwothirtyfive}{\textsc{Qwen3-235B}\xspace}
\newcommand{\gpthundred}{\textsc{gpt-oss-120b}\xspace}
\newcommand{\gpttwenty}{\textsc{gpt-oss-20b}\xspace}

\newcommand{\gemmaone}{\textsc{Gemma3 1B Instruct}\xspace}
\newcommand{\gemmafour}{\textsc{Gemma3 4B Instruct}\xspace}
\newcommand{\gemmatwelve}{\textsc{Gemma3 12B Instruct}\xspace}

\newcommand{\granitehmicro}{\textsc{granite-4.0-h-micro}\xspace}
\newcommand{\granitehsmall}{\textsc{granite-4.0-h-small}\xspace}
\newcommand{\granitehtiny}{\textsc{granite-4.0-h-tiny}\xspace}

\newcommand{\ipw}{\textsc{IPW}\xspace}

\title{Intelligence per Watt:\\{\mdseries Measuring Intelligence Efficiency of Local AI}}

\author{%
  Jon~Saad-Falcon$^{*1}$ \And
  Avanika~Narayan$^{*1}$ \And
  Hakki~Orhun~Akengin$^{1}$ \\
  J.~Wes~Griffin$^{1}$ \And
  Herumb~Shandilya$^{1}$ \And
  Adrian~Gamarra~Lafuente$^{1}$ \\
  Medhya~Goel$^{1}$ \And
  Rebecca~Joseph$^{1}$ \And
  Shlok~Natarajan$^{1}$ \\
  Etash~Kumar~Guha$^{1}$ \And
  Shang~Zhu$^{2}$ \And
  Ben~Athiwaratkun$^{2}$ \\
  John~Hennessy$^{1}$ \And
  Azalia~Mirhoseini$^{1}$ \And
  Christopher~R\'e$^{1}$%
}
\affiliations{%
  $^{*}$\,Equal contribution. \quad
  $^{1}$\,Stanford University \quad
  $^{2}$\,Together AI%
}
\code{https://github.com/HazyResearch/intelligence-per-watt}
\website{https://hazyresearch.stanford.edu/intelligence-per-watt/}
\date{}

\heroabstract{%
Large language model (LLM) queries are predominantly processed by frontier models in centralized cloud infrastructure. Rapidly growing demand strains this paradigm, and cloud providers struggle to scale infrastructure at pace. Two advances create an opportunity to rethink this paradigm: small, local LMs ($\leq$ 20B active parameters) now achieve competitive performance to frontier models on many tasks, and local accelerators (e.g., Apple M4 Max) can host these models at interactive latencies. This raises the question: \textit{can local inference viably redistribute demand from centralized infrastructure?} Answering this requires measuring both whether local LMs can accurately answer real-world queries \textit{and} whether they can do so efficiently enough to be practical on power-constrained devices (i.e., laptops). We propose \textit{intelligence per watt} (\textsc{IPW}), task accuracy divided by unit of power, as a unified metric for assessing both the capability and efficiency of local inference across model-accelerator configurations.
We conduct a large-scale empirical study across 20+ state-of-the-art local LMs, 8 hardware accelerators (local and cloud), and a representative subset of LLM traffic: 1M real-world single-turn chat and reasoning queries.
For each query, we measure accuracy (local LM win rate against frontier models), energy consumption, latency, and power. Our analysis reveals three key findings.
\textbf{First}, local LMs can successfully answer \textbf{$88.7\%$} of single-turn chat and reasoning queries with accuracy varying by domain.
\textbf{Second}, longitudinal analysis from 2023-2025 shows progress in local inference viability: \textsc{IPW} improved \textbf{$5.3\times$}, driven by both algorithmic advances and accelerator improvements, with locally-serviceable query coverage increasing from $23.2\%$ to $71.3\%$.
\textbf{Third}, local accelerators achieve at least \textbf{$1.4\times$} lower \textsc{IPW} than cloud accelerators running identical models, revealing significant headroom for local accelerator optimization.
These findings demonstrate that local inference can meaningfully redistribute demand from centralized infrastructure for a substantial subset of queries, with \textsc{IPW} serving as the critical metric for tracking this transition.%
}

\begin{document}

\maketitle

\section{Introduction}
\label{sec:introduction}

Large language model (LLM) queries are predominantly processed by frontier models deployed in centralized cloud infrastructure~\citep{openai2025stargate, alvarezmarsal2025rethinking}.
This centralized approach faces mounting resource constraints as inference workloads scale from billions to trillions of queries daily~\citep{alvarezmarsal2025rethinking}. History suggests an alternative path forward.
From 1946-2009, computing efficiency (performance-per-watt) doubled every 1.5 years~\citep{koomey2010implications}, enabling a redistribution of computing workloads from data center mainframes to personal computers.
This transition occurred when efficiency improvements enabled computing to meet user needs within personal device power constraints, not when PCs surpassed mainframes in raw performance.

Three converging trends suggest a similar inflection point may be emerging for LLM inference.
First, recent advances have produced \textit{local LMs}: small models ($\leq20B$ active parameters) such as \textsc{Qwen3}~\citep{qwen3technicalreport}, \textsc{Llama3.1}~\citep{grattafiori2024llama3herdmodels}, and \textsc{gpt-oss}~\citep{agarwal2025gptoss} that achieve competitive performance on many benchmarks while requiring less energy and compute than larger, frontier models~\citep{agarwal2025gptoss}.
Second, local accelerators (e.g., Apple M4 Max, AMD Ryzen AI) now have sufficient memory capacity and compute throughput to host these models with interactive latencies~\citep{apple2024m4maxspecs}.
Third, a wave of open-source personal-AI agent stacks designed for on-device execution (e.g., \textsc{OpenClaw}~\citep{steinberger2025openclaw}, \textsc{Hermes Agent}~\citep{nousresearch2025hermes}, \textsc{OpenJarvis}~\citep{saadfalcon2026openjarvis}), \textsc{PicoClaw}~\citep{sipeed2026picoclaw}, and \textsc{ZeroClaw}~\citep{zeroclaw2026} has emerged, reflecting growing interest in local-first system design.
This raises the question: \textit{Can local inference viably redistribute demand from centralized infrastructure?}

\begin{figure*}[t]%
    \centering
    \includegraphics[width=1.0\textwidth]{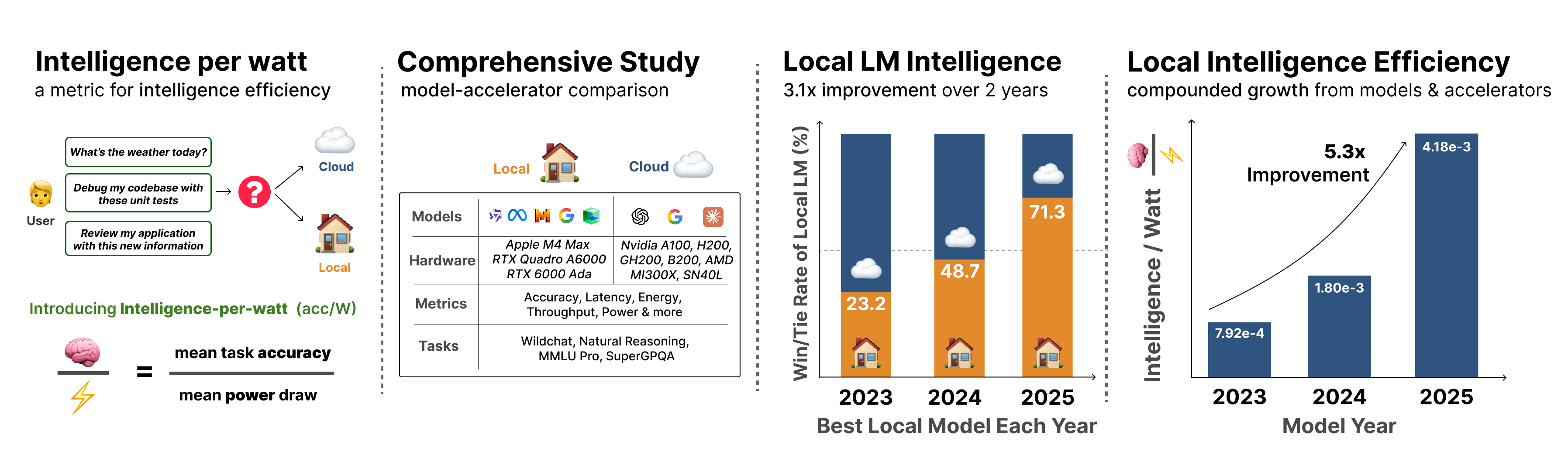}
 \caption{\textbf{Intelligence per Watt: A Study of Local Intelligence Efficiency.}
    We present the \textit{first systematic study of local AI inference efficiency} across models, hardware, and real-world workloads.
    \textbf{(Left)} \textit{Intelligence efficiency is defined as task accuracy per unit of power}, capturing both capabilities delivered and energy consumed.
    \textbf{(Left-Middle)} We conduct \textit{comprehensive performance profiling} across 20+ state-of-the-art local LMs ($\leq20B$ active parameters), diverse hardware accelerators (\textsc{Apple}, \textsc{NVIDIA}, \textsc{AMD}), multiple performance metrics, and 1M+ real-world queries spanning chat and reasoning tasks.
    \textbf{(Right-Middle)} \textit{Local LM capabilities are improving rapidly}: win/tie rate versus frontier models increases from $23.2\%$ (2023) to $71.3\%$ (2025), a $3.1\times$ improvement in accuracy, demonstrating that local models can accurately handle significant portions of single-turn chat and reasoning queries.
    \textbf{(Right)} \textit{Intelligence per watt improves $5.3\times$ from 2023--2025}, driven by advances in both model architectures and hardware accelerators, with local accelerators showing $1.5\times$ efficiency headroom compared to enterprise-grade systems.
}
    \label{fig:trafficbench_overview}
\end{figure*}%

Answering this requires measuring two factors: the \textit{capability} of local LMs to accurately respond to a subset of real-world queries, and the \textit{efficiency} with which local accelerators convert power into useful computation.
To assess this, we need a unified metric that captures both the intelligence delivered (model capability) and the energy required (accelerator efficiency).
We introduce \textit{intelligence per watt} (\ipw): task accuracy per unit of power consumption.
\ipw directly measures the fundamental tradeoff facing local inference: achieving sufficient task performance within constrained power budgets.
This metric enables systematic comparison across model-accelerator configurations and quantifies efficiency gains from model architecture innovations~\citep{qwen3technicalreport, agarwal2025gptoss, gemmateam2025gemma3technicalreport, granite2025}, post-training techniques~\citep{hinton2015distilling, ouyang2022training, shao2024deepseekmath, dettmers2022llm}, and accelerator improvements~\citep{mediangroup_numbers_2019, nvidia_datacenter_gpu, amd_accelerator_specs}.

To evaluate the viability of local inference and measure progress in \ipw, we conduct a large-scale empirical study addressing three questions:
\begin{itemize}
    \item \textbf{Q1:} What fraction of current inference queries can be solved by local LMs on local accelerators, and how has this changed over time?
    \item \textbf{Q2:} How has intelligence per watt improved across successive generations of local models and accelerators, and what are the relative contributions of model versus accelerator advances?
    \item \textbf{Q3:} What resource savings (e.g.\ compute, energy, dollar cost) are possible by distributing workloads across local and cloud infrastructure?
\end{itemize}

Our study evaluates 20+ local LMs across 8 hardware accelerators on 1M queries spanning naturalistic user conversations~\citep{deng2024wildvisopensourcevisualizer}, general reasoning tasks~\citep{yuan2025naturalreasoningreasoningwild28m}, and standardized benchmarks measuring knowledge breadth (\mmlupro{}~\citep{wang2024mmluprorobustchallengingmultitask}) and expert-level reasoning (\supergpqa{}~\citep{pteam2025supergpqascalingllmevaluation}). We focus on single-turn interactions because they constitute a substantial portion of LLM usage~\citep{deng2023chatgpt, wang2024mint, webfx2025chatgpt}.\footnote{We report a multi-turn extension on GAIA and TerminalBenchV2 in App.~\ref{app:multi_turn} confirming the qualitative patterns generalize.}
We compare state-of-the-art local LMs from October 2025 (\textsc{Qwen3}, \textsc{gpt-oss}, \textsc{Gemma3}, \textsc{IBM Granite4}) alongside 2023-2024 models (\textsc{Mixtral-8x7B}, \textsc{Llama-3.1-8B}) on \textsc{NVIDIA}, \textsc{AMD}, and \textsc{Apple} accelerators, measuring accuracy, latency, energy, compute, cost, and memory per query (Section~\ref{sec:tbench_overview}). We release our hardware-agnostic profiling harness to support reproducible efficiency benchmarking.

Our work makes three primary contributions.
\textbf{(1)} We introduce intelligence per watt as a unified metric for evaluating local inference viability, and conduct the first large-scale empirical study measuring its evolution across 1M+ queries, 20+ models, and 8 hardware accelerators spanning 2023-2025.
\textbf{(2)} \textit{(Q1, Q2)} We demonstrate that $88.7\%$ of single-turn chat and reasoning queries can be successfully handled by small local models (with coverage varying by domain), and that IPW has improved $5.3\times$ over two years through compounding model ($3.1\times$) and hardware ($1.7\times$) advances.\footnote{The corresponding per-joule decomposition (Figure~\ref{fig:intelligence_efficiency_trend_bar_plot}) yields $18.0\times$ overall, $3.1\times$ model and $5.9\times$ hardware, since hardware progress disproportionately reduces latency. Both \textsc{Mixtral-8x7B} (47B total, ${\sim}12.9$B active per token) and \gpthundred{} (120B total, $\leq 20$B active per token) are mixture-of-experts models; our $\leq 20$B threshold refers to \textit{active} parameters per forward pass, which determine per-query power and latency. Total parameter count governs storage: at FP4, \textsc{Mixtral-8x7B} fits within 24~GB GDDR6 (Quadro RTX 6000), and \gpthundred{} fits within 128~GB unified memory (Apple M4 Max).}
\textbf{(3)} \textit{(Q3)} We show that hybrid local-cloud routing yields $60$--$80\%$ reductions in energy, compute, and cost compared to a batched cloud baseline; even an $80\%$-accurate router (a realistic target) captures $\sim$$80\%$ of oracle gains while maintaining answer quality.
Together, these findings establish local inference as a practical complement to centralized infrastructure whose viability continues expanding.

\section{Preliminaries}
\label{sec:preliminaries}
We formalize local and cloud inference infrastructure and introduce metrics for measuring intelligence efficiency.

\textbf{Inference Infrastructure: Queries, Models and Accelerators.} We consider an inference infrastructure serving a stream of user queries $\mathcal{Q} = \{q_1, q_2, \ldots, q_n\}$, where each query $q_i$ represents a user-generated request (e.g., chat messages, reasoning tasks).
Let $\mathcal{M}_{\text{local}} = \{m_1, \ldots, m_k\}$ denote a set of \textit{local LMs} with $\leq20B$ active parameters each, and $\mathcal{M}_{\text{cloud}} = \{M_1, \ldots, M_\ell\}$ denote \textit{frontier LMs} with $\geq100B$ parameters.
Similarly, let $\mathcal{H}_{\text{local}}$ represent \textit{local accelerators} (e.g., \textsc{Apple M4}, \textsc{AMD Ryzen}) and $\mathcal{H}_{\text{cloud}}$ represent \textit{cloud accelerators} (e.g., \textsc{NVIDIA H200}, \textsc{AMD MI300X}). We define $\mathcal{M}_{\text{local}}$ by active parameters rather than total parameters because per-query inference efficiency depends on parameters touched per forward pass; total parameters instead govern storage, a separate constraint we verify per accelerator.

\textbf{Inference Serving: Local and Cloud.}
We distinguish between two inference paradigms: \textit{local inference}, where queries are processed by models $m \in \mathcal{M}_{\text{local}}$ on accelerator $h \in \mathcal{H}_{\text{local}}$, and \textit{cloud inference}, where queries are processed by models $M \in \mathcal{M}_{\text{cloud}}$ on accelerator $H \in \mathcal{H}_{\text{cloud}}$. A \textit{routing function} $r: \mathcal{Q} \rightarrow \mathcal{M}_{\text{local}} \cup \mathcal{M}_{\text{cloud}}$ assigns each query to either a local model (up to 20B active parameters) or a cloud model (at least 100B parameters).

\textbf{Intelligence Efficiency Metrics.}
We introduce a family of metrics to quantify how efficiently inference systems convert energy into useful computation.
For a model-accelerator pair $(m, h)$, let $\text{acc}(m, q)$ denote the accuracy of model $m$ on query $q$, $\text{ppl}(m, q)$ denote the perplexity, $P(m, h, q)$ denote the average power consumption (in watts) during inference for query $q$, and $\tau(m, h, q)$ denote the total latency (in seconds) for generating the response, including both prefill and decoding phases.

We define four complementary efficiency metrics:

\textbf{Power-based metrics} measure efficiency relative to instantaneous power draw:
\begin{itemize}
\item \textit{Accuracy per watt}: $\text{APW}(m, h) = \frac{\mathbb{E}_{q \sim \mathcal{Q}}[\text{acc}(m, q)]}{\mathbb{E}_{q \sim \mathcal{Q}}[P(m, h, q)]}$
\item \textit{Perplexity per watt}:

$\text{PPW}(m, h) = \frac{1}{\mathbb{E}_{q \sim \mathcal{Q}}[\text{ppl}(m, q)] \cdot \mathbb{E}_{q \sim \mathcal{Q}}[P(m, h, q)]}$

\end{itemize}

\textbf{Energy-based metrics} measure efficiency relative to total energy consumed per query:
\begin{itemize}
\item \textit{Accuracy per joule}:

$\text{APJ}(m, h) = \frac{\mathbb{E}_{q \sim \mathcal{Q}}[\text{acc}(m, q)]}{\mathbb{E}_{q \sim \mathcal{Q}}[P(m, h, q) \cdot \tau(m, h, q)]}$

\item \textit{Perplexity per joule}:

$\text{PPJ}(m, h) = \frac{1}{\mathbb{E}_{q \sim \mathcal{Q}}[\text{ppl}(m, q)] \cdot \mathbb{E}_{q \sim \mathcal{Q}}[P(m, h, q) \cdot \tau(m, h, q)]}$

\end{itemize}

where $P(m, h, q) \cdot \tau(m, h, q)$ represents the energy consumption (in joules) for processing query $q$.

Power-based metrics (\text{APW}, \text{PPW}) capture the instantaneous efficiency of the inference system, reflecting the hardware's ability to deliver performance at a given power draw.
Energy-based metrics (\text{APJ}, \text{PPJ}) capture the total efficiency per query, accounting for both power consumption and generation latency.
Together, these metrics provide a comprehensive view of inference efficiency: \textit{intelligence per watt} quantifies the steady-state efficiency of model-accelerator pairs, while \textit{intelligence per joule} quantifies the end-to-end efficiency from a user's perspective, including the time cost of generation.
We report results across all four metrics throughout the paper (\autoref{tab:longitudinal_trends_for_local_cloud}, Figures~\ref{fig:intelligence_efficiency_trend_bar_plot} and~\ref{fig:perplexity_and_accuracy_trends}, Tables~\ref{tab:apple_m4_max_vs_nvidia_b200_for_power}--\ref{tab:apple_m4_max_vs_nvidia_b200_for_energy}) so that conclusions are robust to the choice of formulation; relative rankings and qualitative trends are preserved across IPW, IPJ, PPW, and PPJ, while absolute rates of change differ in informative ways (e.g., per-joule gains exceed per-watt gains because hardware progress reduces both power draw and generation latency).
Throughout the main text we use \textit{accuracy} as a binary indicator $\text{acc}(m, q) \in \{0, 1\}$: for benchmarks with ground-truth answers (\mmlupro{}, \supergpqa{}, \nr{}) we use exact-match correctness, while for open-ended chat queries (\wildchat{}) we follow established practice in chat evaluation~\citep{chiang2024chatbot} and define $\text{acc}(m, q) = 1$ whenever the LLM-judge verdict is \texttt{[[A>B]]}, \texttt{[[A>>B]]}, or \texttt{[[A=B]]} (i.e., the local model wins or ties against the frontier reference).

\section{Dataset and Profiling Harness}
\label{sec:dataset_and_profiling}
\label{sec:tbench_overview}
In this section, we provide details on the dataset selection and profiling harness.

\subsection{Dataset Selection}

\textbf{Query Curation} We curate over $1M$ queries across four complementary benchmarks designed to measure both naturalistic deployment scenarios and controlled capability assessment.
To ensure our findings about local inference efficiency generalize across task distributions, we combine naturalistic queries that reflect real-world LLM usage patterns with standardized benchmarks that enable systematic evaluation of knowledge breadth and reasoning capabilities across diverse domains.

For \textit{naturalistic chat tasks}, we source queries from  \wildchat~\citep{deng2024wildvisopensourcevisualizer}: a dataset of $1M$ real ChatGPT prompts, spanning 1 month of user traffic.
For \textit{general reasoning tasks}, we source queries from  \nr~\citep{yuan2025naturalreasoningreasoningwild28m}, which provides approximately 1.2 million reasoning-focused queries spanning diverse domains including mathematics, physics, and chemistry.
For \textit{standardized knowledge evaluation}, we use \mmlupro{}~\citep{wang2024mmluprorobustchallengingmultitask}: an enhanced version of MMLU with increased difficulty (10 vs. 4 answer choices) and improved robustness to prompt variations, measuring multi-domain knowledge understanding.
For \textit{expert-level reasoning across specialized disciplines}, we evaluate on \supergpqa{}~\citep{pteam2025supergpqascalingllmevaluation}: a comprehensive benchmark spanning 285 graduate-level disciplines with emphasis on technical domains and specialized fields underrepresented in typical evaluations (e.g., light industry, agriculture, service sciences).

We perform robust data cleaning and filtering (see App.~\ref{app:data_generation_procedure}) on each dataset before sampling queries: $500K$ from \wildchat, $500K$ from \nr, $12K$ from \mmlupro{}, and $26.5K$ from \supergpqa{} (see Table~\ref{tab:tbench_overview}).
Furthermore, we use \textsc{gpt-4o-mini} to annotate each query with a category from the Anthropic Economic Index~\citep{handa2025economictasksperformedai}, which maps AI queries to occupations in the U.S. Department of Labor's O*NET.
We consider 22 categories, spanning ``Architecture and Engineering'' to ``Healthcare Support'' (full list and category breakdown in App.~\ref{app:data_generation_procedure}, Table~\ref{tab:anthropic_economic_index_categories}).

\begin{table}[h]
\centering
\scriptsize
\begin{minipage}[t]{0.40\linewidth}
\centering
\vspace{0pt}
\begin{tabular}{@{}llr@{}}
\toprule
\textbf{Dataset Origin} & \textbf{Category} & \textbf{$|N|$} \\
\midrule
\wildchat{} & Chat & 500K \\
\nr{} & Reasoning & 500K \\
\mmlupro{} & Knowledge & 12K \\
\supergpqa{} & Grad. Reasoning & 26.5K \\
\bottomrule
\end{tabular}
\end{minipage}%
\hfill%
\begin{minipage}[t]{0.58\linewidth}
\centering
\vspace{0pt}
\begin{tabular}{@{}lp{0.7\linewidth}@{}}
\toprule
\textbf{Category} & \textbf{Items} \\
\midrule
Model Families & \textsc{Qwen3}, \textsc{gpt-oss}, \textsc{Gemma}, \textsc{IBM Granite 4.0} \\
\midrule
Accelerators & \textsc{NVIDIA A100}, \textsc{H200}, \textsc{GH200}, \textsc{B200}, \textsc{Quadro RTX 6000}, \textsc{RTX 6000 Ada}, \textsc{AMD MI300X}, \textsc{Apple M4 Max}, \textsc{Sambanova SN40L} \\
\bottomrule
\end{tabular}
\end{minipage}
\caption{\textbf{Dataset Overview.} \textbf{(Left)} Query composition with sizes. \textbf{(Right)} Models and Accelerators.}
\label{tab:tbench_overview}
\end{table}

\textbf{Hardware Accelerators} We profile diverse accelerators spanning local, workstation, and datacenter tiers: the \textsc{NVIDIA A100 40 GB SXM4 (Ampere)}~\citep{nvidia2021a100datasheet}, \textsc{NVIDIA H200 SXM (Hopper)}~\citep{nvidia2024h200datasheet}, \textsc{NVIDIA GH200 Grace Hopper Superchip}~\citep{nvidia2024gh200datasheet}, \textsc{NVIDIA B200 (Blackwell)}~\citep{nvidia2025b200datasheet}, \textsc{NVIDIA Quadro RTX 6000}~\citep{nvidia_quadro_rtx6000_2019}, NVIDIA RTX 6000 Ada~\citep{nvidia_rtx6000_ada_2023}, \textsc{AMD Instinct MI300X (CDNA 3, OAM)}~\citep{amd2023mi300x}, SambaNova SN40L~\citep{sambanova2022datasheet} and \textsc{Apple Mac Studio (M4 Max)}~\citep{apple2024m4maxspecs}.
We additionally evaluate a smartphone-class accelerator (\textsc{Apple A18 Pro} on iPhone 16 Pro) in App.~\ref{app:smartphone_experiments}.
These systems were chosen because of their different memory capacities (ranging from 40 GB to 768 GB), memory bandwidth (from 546 GB/s to 8 TB/s), and power consumption (145W to 1000W) (see Table~\ref{tab:accelerator_details} for more details).

\textbf{Models} We collect model generations over the \textsc{Qwen3}~\citep{qwen3technicalreport}, \textsc{gpt-oss}~\citep{agarwal2025gptoss}, \textsc{Gemma3}~\citep{gemmateam2025gemma3technicalreport}, and \textsc{IBM Granite 4.0}~\citep{granite2025} families. For \textsc{Qwen3}, we use \qwenfour, \qweneight, \qwenfourteen, \qwenthirtytwo, and \qwentwothirtyfive.
For \textsc{GPT-OSS}, we consider the \gpttwenty and \gpthundred models.
For the \textsc{Gemma3} family, we use \gemmaone{}, \gemmafour{}, and \gemmatwelve{} models.
For \textsc{IBM Granite 4.0}, we use \granitehmicro{}, \granitehtiny{}, and \granitehsmall{} models.
We evaluate state-of-the-art cloud models as of October 2025, including \textsc{Claude Sonnet 4.5} \citep{anthropic2025claude45}, \textsc{Gemini 2.5 Pro} \citep{comanici2025gemini25pushingfrontier}, and \textsc{GPT-5} (2025-08-07) \citep{openai2025gpt5}.
For our longitudinal analysis, we evaluate \textsc{Mixtral-8x7B}~\citep{jiang2024mixtral} and \textsc{Llama3.1-8B}~\citep{grattafiori2024llama3herdmodels}.
For each model, we generate responses across all dataset queries on each of the hardware backends.
Full details of inference hyperparameters can be found in App.~\ref{app:data_generation_procedure}.

\textbf{Metrics} For each \texttt{(query, model, hardware)} triple we collect accuracy plus efficiency metrics: latency, throughput, time-to-first-token (TTFT), and more (see Table~\ref{tab:full_metrics_profile})
We use LLM-as-a-judge (prompts in App.~\ref{app:metrics}) to score generated responses against reference answers.
For \wildchat, reference answers are responses from \qwentwothirtyfive, the SOTA open-source model on LMArena (as of August 2025)~\citep{chiang2024chatbot}.
For \nr, \mmlupro{}, and \supergpqa{}, we use the provided ground truth answers from each benchmark.

\subsection{Profiling Harness}
We develop an end-to-end, cross-platform profiling harness for inference workloads that ensures reproducible results and easily accommodates new models, tasks, and hardware backends. It comprises three components (distributed multi-GPU inference, response evaluation, and system-level telemetry collection) and currently supports \textsc{NVIDIA}, \textsc{macOS} (Apple Silicon), and \textsc{AMD} systems. Given a dataset, model, and backend, the harness orchestrates inference over all input queries, evaluates outputs (via exact match or LLM-as-a-judge), and records detailed telemetry: latency, throughput, time-to-first-token (TTFT), energy consumption, and more (Table~\ref{tab:full_metrics_profile}). Telemetry is collected via vendor APIs, synchronized at nanosecond resolution, and normalized to common units (watts, joules, megabytes). For energy measurements, we follow standard practices~\citep{samsi2023words, fernandez2025energy, wilkins2024hybrid}. On \textsc{NVIDIA} systems we query NVML for per-device power, energy, memory usage, and temperature (accelerator-only scope); on \textsc{AMD} systems we query ROCm SMI for power, temperature, and VRAM usage (accelerator-only scope); on \textsc{macOS} systems we extract GPU power from \texttt{powermetrics} (\texttt{processor\_power.actual} on Apple Silicon, isolating the GPU subsystem rather than full SoC package power) so that all per-query power measurements correspond to the AI-accelerator subsystem on each platform. In all cases, we compute energy via numerical integration over time and sample at 50\,ms intervals, providing higher temporal resolution than prior work (100\,ms~\citep{samsi2023words} or 15\,s~\citep{fernandez2025energy}). For multi-GPU configurations, we aggregate energy from each GPU individually rather than extrapolating from a single device~\citep{samsi2023words}. We use a custom harness rather than \textsc{CodeCarbon}~\citep{codecarbon2024}, which similarly relies on NVML and RAPL, primarily for two reasons: (i) \textsc{CodeCarbon}'s default sampling cadence (15\,s) is too coarse for fine-grained per-query attribution on short generations, and (ii) our cross-platform requirements include AMD (ROCm SMI) and Apple Silicon (\texttt{powermetrics}) telemetry that \textsc{CodeCarbon} does not natively support. Software-based power measurements can introduce inaccuracies of 10--15\%, with variations distributed across different hardware components due to architectural differences in workloads between CPUs, GPUs, and NPUs~\citep{yang2023part}. Even hardware wattage meters may fall short for milliwatt-level precision, though our approach aligns with established practices and provides consistent relative comparisons across configurations. Full implementation details are provided in App.~\ref{app:telemetry_collection}.

\section{Intelligence Efficiency Study}
\label{sec:tbench_study}

\begin{figure*}[t]
    \centering
    \includegraphics[width=\linewidth]{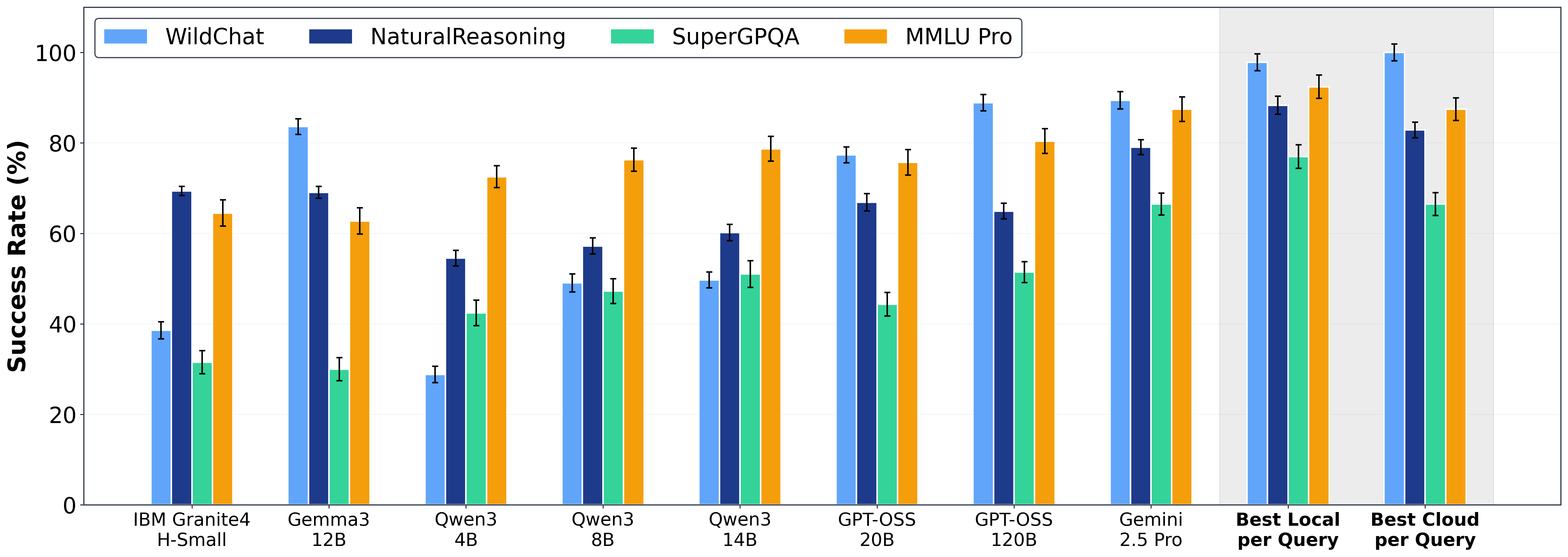}
    \caption{\textbf{Local Models Rival Cloud Models Across Diverse Benchmarks:}
        Individual model performance scales with size, ranging from 31.5--69.4\% for \textsc{IBM~Granite4-H-Small}, 30.0--83.6\% for \textsc{Gemma3-12B}, 51.5--80.4\% for \textsc{GPT-OSS-120B}, and 66.5--89.5\% for \textsc{Gemini~2.5~Pro}.
        Local routing (best local LM per query) achieves 97.8\%, 88.3\%, 77.0\%, and 92.4\% on \wildchat, \nr, \supergpqa{}, and \mmlupro{} respectively, surpassing cloud routing (100\%, 82.9\%, 66.5\%, 87.4\%) on three of four benchmarks.
    }
    \label{fig:query_coverage}
\end{figure*}

We investigate whether recent advances in local LMs and local accelerators enable local inference to viably complement centralized cloud infrastructure by handling a substantial fraction of inference queries. Using our curated dataset, we examine three interconnected questions: \textbf{(1)} the extent to which current workloads can be handled locally (Section~\ref{sec:local_lms_on_inference_workloads}), \textbf{(2)} how intelligence efficiency has evolved from 2023--2025 (Section~\ref{sec:intelligence_efficiency_trends}), and \textbf{(3)} what gains query routing across local and cloud models can deliver in practice (Section~\ref{sec:efficiency_gains_from_routing}). We use single-query inference ($\text{bs}=1$) to isolate intrinsic model-accelerator efficiency from system-level scheduling and follow standard local-inference benchmarking practice~\citep{hao2023reaching}; the §4.3 routing simulation is the exception, running the cloud baseline at bs=16 to reflect production serving.

\subsection{Can Local Models and Accelerators Handle Current Inference Workloads? \textit{(Q1)}}
\label{sec:local_lms_on_inference_workloads}

We measure \textit{query coverage} (the percentage of dataset queries answered correctly) across three configurations (Figure~\ref{fig:query_coverage}): individual local LMs, the best-of-local ensemble (routing to the best local LM), and the best-of-cloud baseline (routing to the best frontier model). Our findings are as follows:

\textbf{Local LM coverage increases with scale and time.}
Across \wildchat{}, \nr{}, \supergpqa{}, and \mmlupro{}, individual model coverage ranges from $49.6\%$ for \textsc{Qwen3-4B}, on average, to $71.4\%$ for \gpthundred{}, with consistent improvements at each scale point: \textsc{Qwen3-8B} achieves $57.5\%$ and \textsc{Qwen3-14B} reaches $60.0\%$.
Coverage has improved substantially from 2023 to 2025 (Figure~\ref{fig:chat_and_reasoning_trends}): the best local LMs achieved a $32.2\%$ relative improvement on chat queries and a $50.1\%$ relative improvement on reasoning queries over this period.
While improvements from 2023-2025 are relatively uniform across difficulty levels for chat tasks, reasoning tasks show markedly slower progress on the hardest problems (see App~\ref{app:tbench_experiments}).
These results demonstrate that larger local LMs can handle progressively more queries without requiring cloud infrastructure, with the best individual local LM (\gpthundred) successfully answering almost three-fourths of the single-turn chat and reasoning queries studied.

\textbf{Model diversity substantially improves coverage.}
Routing queries to the most appropriate local LM rather than using a single model achieves $88.7\%$ overall coverage, a $28.8$ percentage point improvement over \textsc{Qwen3-14B} and $16.3$ percentage points over individual \gpthundred performance, on average.
This gap between individual models and best-of-local demonstrates that architectural, pretraining, and post-training diversity captures complementary capabilities: different models excel on different query types, and intelligent routing can exploit these complementary strengths.
On reasoning benchmarks, best-of-local even surpasses best-of-cloud (\autoref{fig:query_coverage}); this is a best-of-$N$ selection effect, since best-of-local selects from $20+$ diverse local models while best-of-cloud selects from three frontier models, and a sufficiently diverse local ensemble can therefore exceed any single frontier model on subsets of queries where local strengths are complementary.

\textbf{Chat queries are more amenable to local processing than reasoning queries.}
The best local LM achieves $88.9\%$ coverage on \wildchat{} versus $64.9\%$ on \nr, a $24.0$ pp gap consistent with findings that $77\%$ of real-world ChatGPT queries involve practical guidance, information seeking, or writing~\citep{chatterji2025chatgpt}. These tasks are well-suited to local models, while reasoning-intensive queries more often require frontier capabilities for technical domains (Architecture, Engineering, Life \& Physical Science; see Figure \ref{fig:routing_share_by_domain} in App. \ref{app:dataset}). Even on \nr{}/\supergpqa{}/\mmlupro{}, local LMs handle over four-fifths of reasoning queries studied, suggesting significant opportunities for local inference even in technically demanding domains.

\textbf{Evaluation on standardized benchmarks confirms local LM viability across task distributions.}
On \mmlupro (multi-domain knowledge) and \supergpqa (graduate-level reasoning), best-of-local achieves $93.4\%$ and $83.6\%$ coverage respectively (vs.\ $80.4\%$ and $51.5\%$ for the best individual local model), with coverage exceeding $93\%$ for creative/humanities fields but dropping to $60\%$ for technical disciplines like Architecture \& Engineering (Figure~\ref{fig:routing_share_by_domain}), confirming that local LMs handle most conversational and knowledge-recall tasks while complex specialized reasoning still benefits from frontier capabilities.

\textbf{Local accelerator memory capacity is expanding rapidly.}
From 2012 to 2025, local accelerator memory grew $\sim$$126\times$ (Figure~\ref{fig:gpu_frontier} in App.~\ref{app:local_vs_cloud_intelligence_efficiency}); the jump from sub-20\,GB to 200+\,GB through unified-memory architectures like Apple Silicon removes the key constraint that previously forced workloads to cloud infrastructure, enabling the $8$--$20$B-active-parameter models that handle the majority of queries today to run efficiently on local hardware.

\subsection{How Intelligence Efficient is Local Inference? \textit{(Q2)}}
\label{sec:intelligence_efficiency_trends}

\begin{table}[t]
\centering
\scriptsize
\setlength{\tabcolsep}{2pt}
\begin{tabular}{lccc}
\toprule
 & \textbf{2023} & \textbf{2024} & \textbf{2025} \\
\midrule
\makecell{\textbf{SOTA} \\ \textbf{Local Model}} & \href{https://huggingface.co/mistralai/Mixtral-8x7B-v0.1}{Mixtral-8x7B-v0.1} & \href{https://huggingface.co/meta-llama/Llama-3.1-8B-Instruct}{Llama-3.1-8B-Instruct} & \href{https://huggingface.co/openai/gpt-oss-120b}{GPT-OSS-120B} \\ \midrule
\makecell{\textbf{SOTA} \\ \textbf{Accelerator}} & \makecell{{NVIDIA Quadro} \\ {RTX 6000}}  & \makecell{{NVIDIA RTX} \\ {6000 Ada}} & \makecell{{Apple} \\ {M4 Max}}\\ \midrule
\makecell{\textbf{Success Rate}} & $23.2 \pm 1.9\%$ & $48.7 \pm 2.7\%$ & $71.3 \pm 2.2\%$ \\ \midrule
\makecell{\textbf{Intelligence} \\ \textbf{per Watt}} & \makecell{$(7.92 \pm 0.32)$ \\ $\times 10^{-4}$} & \makecell{$(1.80 \pm 0.21)$ \\ $\times 10^{-3}$} & \makecell{$(4.18 \pm 0.53)$ \\ $\times 10^{-3}$} \\ \midrule
\makecell{\textbf{YoY Efficiency} \\ \textbf{Gain}} & --- & 2.27× & 2.32× \\
\bottomrule
\end{tabular}
\caption{\textbf{Increase in Intelligence per Watt for Local LMs}: Accuracy per watt has improved over $5\times$ in two years, driven by advances in both model architectures (from \textsc{Mixtral-8x7B} to \textsc{GPT-OSS-120B}) and accelerator hardware (from NVIDIA Quadro RTX 6000 to Apple M4 Max). Values are mean $\pm$ 1-$\sigma$ standard deviation across measurement runs.
}
\label{tab:longitudinal_trends_for_local_cloud}
\end{table}

\textbf{Intelligence efficiency is improving over time}. Table~\ref{tab:longitudinal_trends_for_local_cloud} tracks the evolution of local LM capabilities from 2023 to 2025, measuring the best available local LM ($\leq 20B$ active parameters) paired with state-of-the-art accelerators each year.
On our curated dataset of chat and reasoning queries, accuracy per watt has improved $5.3\times$ over this two-year period: in 2023, \textsc{Mixtral-8x7B-v0.1} on \textsc{NVIDIA Quadro RTX 6000} achieved $7.92 \times 10^{-4}$ accuracy per watt; by 2024, \textsc{Llama-3.1-8B-Instruct} on \textsc{NVIDIA RTX 6000 Ada} reached $1.80 \times 10^{-3}$ (a $2.27\times$ year-over-year gain); and in 2025, \gpthundred on \textsc{Apple M4 Max} achieved $4.18 \times 10^{-3}$ (a $2.32\times$ gain).
Notably, local LM coverage on single-turn chat and reasoning queries has increased in lockstep with efficiency gains: from $23.2\%$ in 2023 to $48.7\%$ in 2024 to $71.3\%$ in 2025.
This progression reflects compounding improvements in both model architectures, which \textit{achieve higher accuracy} through advances in pretraining \citep{chowdhery2022palm, hoffmann2022training, openai2023gpt4, deepseekai2024deepseekv3}, post-training \citep{bai2022constitutional, shao2024deepseekmath, deepseekai2025deepseekr1}, and parameter utilization via mixture-of-experts (MoE) architectures \citep{shazeer2017outrageously, deepseekai2024deepseekv3}, and hardware accelerators, which deliver \textit{more compute (FLOPs) and memory per watt} \citep{nvidia2021a100datasheet, nvidia2024h200datasheet}.

\begin{figure}[!t]
    \centering
    \includegraphics[width=0.9\linewidth]{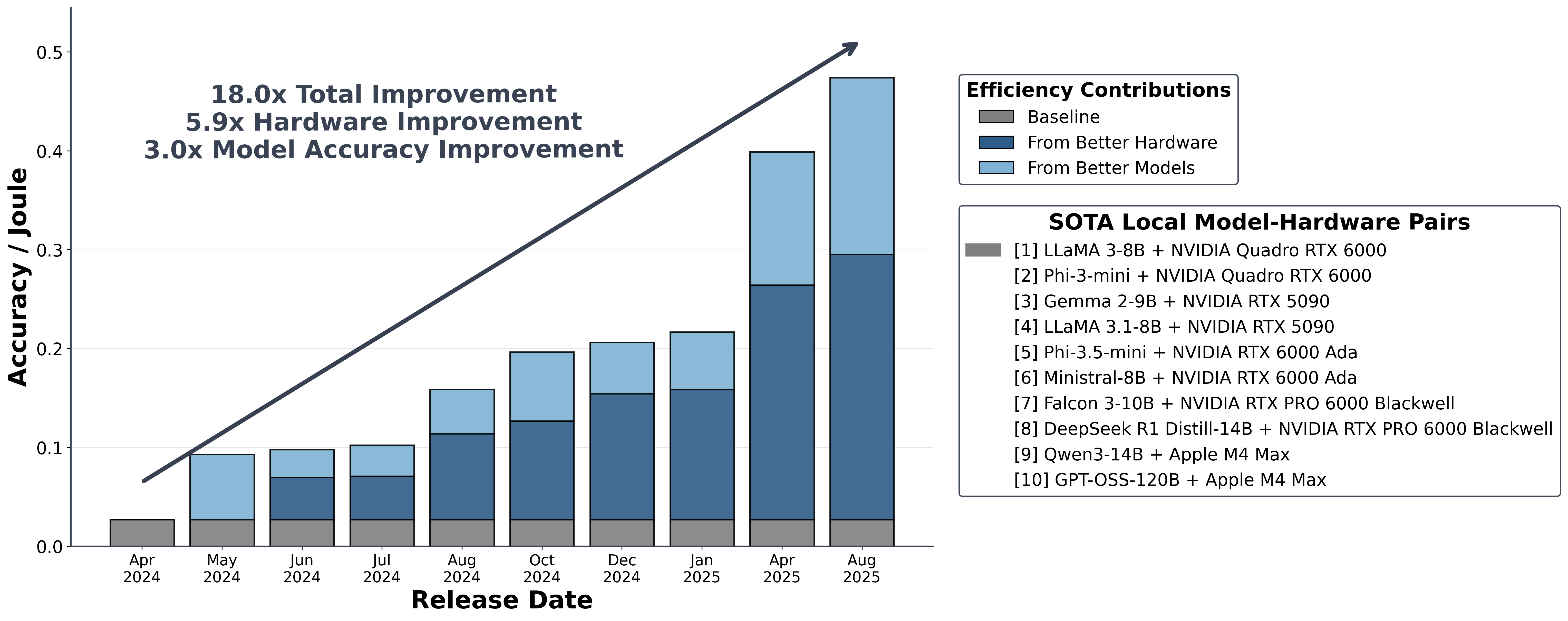}
    \caption{
    \textbf{Increase in Intelligence per Joule for Local LMs and Accelerators}: Efficiency improved $18.0\times$ over 16 months, decomposed into $3.0\times$ from  local LMs and $5.9\times$ from local accelerators.
    }
    \label{fig:intelligence_efficiency_trend_bar_plot}
\end{figure}

The decomposition depends on the metric: under accuracy-per-watt, model
progress contributes 3.1× and hardware 1.7× (Table~\ref{tab:longitudinal_trends_for_local_cloud}),
while under accuracy-per-joule it yields 3.1× and 5.9× respectively
(Figure~\ref{fig:intelligence_efficiency_trend_bar_plot}). Model progress
dominates per-watt efficiency; hardware progress dominates per-joule
efficiency, because newer accelerators (HBM3e bandwidth, dedicated tensor
units) reduce latency as well as power. Both matter — per-watt for
thermally-constrained deployment, per-joule for end-to-end energy budgets —
and Figure~\ref{fig:intelligence_efficiency_trend_bar_plot} shows the per-joule
trend is consistent across nine model families
(\textsc{Llama}, \textsc{Phi}, \textsc{Gemma}, \textsc{Mistral},
\textsc{Falcon}, \textsc{DeepSeek}, \textsc{Qwen}, \textsc{gpt-oss}), with
perplexity-based confirmation in
Figure~\ref{fig:perplexity_and_accuracy_trends} (App.~\ref{app:additional_efficiency_trends}).
Sustained progress in MoE architectures, quantization, and unified-memory
capacity is the precondition for these trends to continue. We explore quantization tradeoffs (App.~\ref{app:model_precision_vs_accuracy}) and serving-stack sensitivity --- batching (App.~\ref{app:batching_ablation}) and framework choice (App.~\ref{app:framework_sensitivity}), finding that FP4 saves 3–3.5$\times$ energy per precision step, cloud bs=$64$ gives 11–20$\times$ higher IPJ than bs=$1$, and IPW rankings across vLLM, SGLang, and llama.cpp are preserved.

\textbf{Local accelerator efficiency has room for improvement}:
While local accelerators enable deployment of capable models outside data centers, cloud-grade hardware maintains a substantial efficiency advantage on the same workloads. Across \textsc{Qwen3} and \textsc{GPT-OSS} variants, the \textsc{NVIDIA B200} achieves $1.40\times$ higher IPW and $1.6$--$2.3\times$ higher IPJ than \textsc{Apple M4 Max}, and \textsc{SambaNova SN40L} achieves up to $1.78\times$ higher IPW and $6.5$--$7.4\times$ higher IPJ (Tables~\ref{tab:apple_m4_max_vs_nvidia_b200_for_power}--\ref{tab:apple_m4_max_vs_nvidia_b200_for_energy} in App.~\ref{app:local_vs_cloud_intelligence_efficiency}). Similar trends are observed on multi-turn agentic workloads (see App. ~\ref{app:multi_turn}). Per-joule gaps widen relative to per-watt gaps because cloud accelerators not only consume less power per unit of accuracy but also complete queries faster. These gaps stem from purpose-built components in enterprise accelerators (HBM3e, dedicated tensor units, optimized memory hierarchies), whereas local accelerators use unified-memory architectures that balance diverse workloads under thermal and power constraints. For these comparisons we use bs = $1$ for both local and cloud, so the gap is intrinsic rather than a batching artifact, revealing substantial headroom for on-device AI components.

\subsection{What Efficiency Gains Can Effective Query Routing Deliver? \textit{(Q3)}}
\label{sec:efficiency_gains_from_routing}

\begin{figure*}[t]
    \centering
    \includegraphics[width=0.9\textwidth]{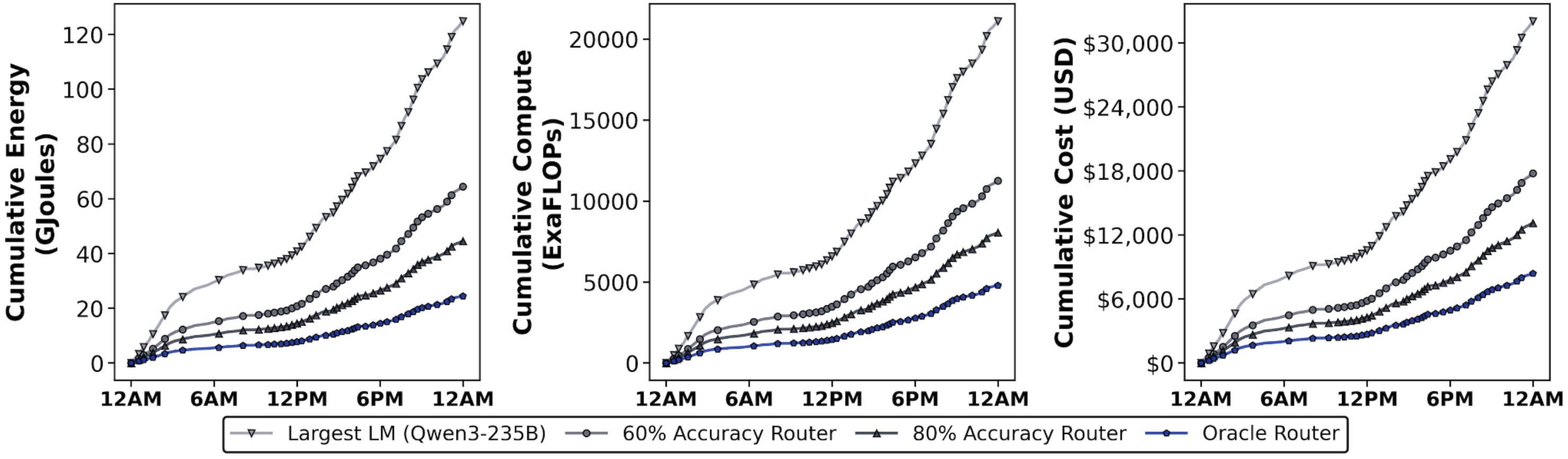}
    \caption{\textbf{Energy, Compute, and Capital Gains from Model Routing}. Cumulative resource consumption over 24 hours and 80.2M LLM queries~\citep{BurstGPT}, routing between $4$ small LMs on Apple M4 Max (bs = $1$) and \qwentwothirtyfive on an H200 (bs = $16$, batched cloud baseline). The 80\%-accurate router achieves 64.3\% energy, 61.8\% compute, and 59.0\% cost savings versus routing all queries to \qwentwothirtyfive, capturing the majority of gains achievable by the Oracle.}
    \label{fig:burst_gpt_plot}
\end{figure*}

We simulate a hybrid local-cloud system serving $80.2$M queries over 24 hours, representative of realistic daily inference workloads~\citep{BurstGPT}. Queries are routed between four small local LMs (\textsc{Qwen3-4B/8B/14B}, \gpttwenty) on \textsc{Apple M4 Max} (bs = $1$) and a frontier model (\textsc{Qwen3-235B}) on \textsc{NVIDIA H200} (bs = $16$, reflecting production serving conditions). Figure~\ref{fig:burst_gpt_plot} compares five strategies: routing all queries to the largest model (baseline), oracle, and realistic routers at $60\%$/$80\%$ routing accuracy; misrouted queries fall back to the cloud model.

\textbf{Oracle routing establishes theoretical upper bounds.}
Assuming perfect query-to-model assignment, oracle routing reduces energy by $80.4\%$, compute by $77.3\%$, and cost by $73.8\%$ versus cloud-only deployment to the largest model (Figure~\ref{fig:burst_gpt_plot}). The $80.7\%$ of queries that local models can handle correctly are assigned to substantially smaller local models, while frontier compute is reserved for the remaining $19.3\%$. The dominant effect is the reduction in model size, not a hardware efficiency advantage. The cloud-only baseline already operates at $\text{bs}=16$, so the reported savings are computed against a batched cloud baseline rather than $\text{bs}=1$ (see App.~\ref{app:batching_ablation}).

\textbf{Practical routers achieve substantial gains without perfect accuracy.}
A router with $80\%$ accuracy (a realistic target, with prior work such as RouteLLM~\citep{ong2025routellm} reporting $70$--$85\%$ on real workloads) captures $\sim$$80\%$ of oracle gains: $64.3\%$ energy, $61.8\%$ compute, and $59.0\%$ cost reduction. Even a $60\%$-accurate router delivers $48.4\%/46.7\%/44.5\%$ savings, with misrouted queries falling back to the frontier model so end-to-end answer quality is maintained. The oracle uses ground-truth correctness as a theoretical upper bound; deployable routers instead use \textit{pre-routing} confidence estimation, and the $60\%/80\%$ scenarios above explicitly model this imperfection. At billions of queries daily, these savings scale linearly to annual energy savings in terawatt-hours. Although local accelerators are less efficient per query (§\ref{sec:intelligence_efficiency_trends}), routing inverts this at the system level: efficient AI infrastructure comes not from local hardware matching cloud, but from routing across both.

\section{Conclusion and Key Takeaways}
\label{sec:conclusion}

We ask whether local inference can viably redistribute demand from centralized cloud infrastructure, and introduce \textit{intelligence per watt} (IPW)---task accuracy per unit of power---as a unified metric for jointly evaluating local LM capability and local accelerator efficiency. We take a first step towards answering this question through a longitudinal study of  $20+$ models, $8$ accelerators, and $1$M queries spanning 2023--2025. We find that $88.7\%$ of single-turn chat and reasoning queries can be handled locally, IPW has improved $5.3\times$ over two years through compounding model ($3.1\times$) and hardware ($1.7\times$) advances, and although cloud accelerators retain a $1.4$--$7.4\times$ per-query efficiency edge (App.~\ref{app:local_vs_cloud_intelligence_efficiency}), hybrid local-cloud routing yields $60$--$80\%$ aggregate energy, compute, and cost reductions at realistic routing accuracy. Our study provides three practical takeaways  (App.~\ref{app:key_takeaways}):  MoE architectures deliver the best IPW on memory-rich local devices, aggressive FP4 quantization beats smaller-but-higher-precision models, and router accuracy past $\sim$80\% matters less than expanding the local-model ensemble.
We release our profiling harness to support periodic re-evaluation as the local-inference ecosystem evolves, and discuss the scope and limitations of our study in App.~\ref{app:limitations_and_broader_impacts}. In future work, we hope to extend IPW characterization to broader workload regimes (i.e., multi-modal inference) and hybrid local-cloud execution patterns, and to push the local AI frontier through model-hardware co-design (App.~\ref{app:future_work}).

\bibliographystyle{plainnat}
\begingroup
\sloppy
\bibliography{example_paper}
\endgroup

\clearpage
\appendix

\section{Related Works}
\label{app:related_works}

Below, we provide an extended treatment of related works.

\paragraph{LLM Routing} A central challenge in local-cloud routing systems is determining which model should handle a given query so as to maximize efficiency. Prior work spans a broad design space, but much of it can be organized around two families of approaches: embedding-based routers~\citep{zhang2025avengers, somerstep2025carrotcostawarerate, chen2024routerdc} and generative/decoder-based routers~\citep{ong2025routellm}. Embedding-based methods rely on encoding queries (and sometimes models) into a vector space and then applying similarity search or lightweight classification. Early work largely adopted binary routing, where queries are directed between just two models. For example, RouteLLM~\citep{ong2025routellm} demonstrated that simple supervised classification can yield up to 85\% cost reduction while maintaining GPT-4-level performance, but this setting was restricted to two-model scenarios. More recent systems generalize routing to multi-model settings: ensemble-style methods such as FrugalGPT \citep{chen2023frugalgpt}, RouterDC \citep{chen2024routerdc}, and Avengers Pro~\citep{zhang2025avengers,zhang2025beyond} show that intelligently combining smaller models can approximate or even surpass larger frontier LMs. Decoder-based methods leverage a small language model to directly generate the routing decision. Causal LLM Routing, suggests that incorporating richer query-model interaction signals via generative modeling or cross-attention can yield more robust routing than static embeddings~\citep{chen2024routerdc}. In this work, we are inspired by these novel approaches to routing, and evaluate their performance in the local-cloud routing setup.

\paragraph{LLM Routing Benchmarks}
Recent work has explored benchmarks for LLM query routing, primarily targeting cost--quality tradeoffs across multiple models. RouterBench~\citep{hu2024routerbench} provides a comprehensive suite of curated academic tasks (~405K samples) to evaluate routing policies along cost--quality Pareto frontiers. RouteLLM~\citep{ong2025routellm} introduces a preference-trained routing framework evaluated on academic benchmarks like MMLU and MT-Bench, with a focus on achieving quality under token cost constraints, though it remains limited to token-level metrics. RouterEval~\citep{huang2025routereval} emphasizes model selection accuracy at scale, compiling over 200M performance records across 8.5K models and 12 benchmarks to study generalization, yet lacks coverage of real-world queries. In contrast, our curated dataset targets routing under naturalistic conditions, leveraging 1M real user queries from \wildchat and \nr. It uniquely supports the exploration of local-cloud routing tradeoffs beyond just cost and quality, to metrics such as latency, energy, memory, throughput, and more, generated on local accelerators and enterprise-grade accelerators. Moreover, in contrast to existing benchmarks, which provide stale performance records limited to models released prior to July 2024, our curated dataset evaluates several state-of-the-art models, including Qwen3~\citep{qwen3technicalreport} and GPT-OSS~\citep{agarwal2025gptoss}, all released after May 2025. To support ongoing benchmarking, we release our efficiency profiling harness, a hardware-agnostic toolkit for generating fresh telemetry and evaluation records as new models become available.

\paragraph{Local--Cloud Inference Systems}
Beyond model selection, recent work explores collaborative inference protocols that split generation between local and cloud models. Minions~\citep{narayan2025minions} proposes a two-stage protocol where a small on-device LM handles lightweight processing and a frontier LM performs high-level reasoning, with an extended version introducing task decomposition and aggregation for improved quality. Such collaborative schemes offer large energy and cost savings but require careful protocol design to avoid performance loss. A parallel line of work centers on speculative decoding, where a small draft model generates candidate continuations that are verified or refined by a larger target LM~\citep{miao2023specinfer,xu2025specee}. These approaches primarily target latency and throughput, particularly in constrained hardware settings, and typically assume that generation will ultimately invoke a large LM. Other hybrid protocols like SLED~\citep{li2025sled} and HAT~\citep{xie2025hat} introduce edge-cloud model partitioning with intermediate state exchange to balance device limitations with quality needs. While these systems explore fine-grained collaboration at the token or layer level, our work investigates the limitations of a coarser-grained alternative: query-level routing across multiple small and large LMs, where we measure not only accuracy and cost, but also latency, memory, and energy across diverse hardware accelerators.

\paragraph{Efficient AI} We are inspired by work on ``Green AI'' which proposes treating energy as a first-class metric alongside accuracy and cost, with calls for standardized reporting and tooling for reproducible accounting of power use and emissions during training and inference \citep{schwartz2020greenai,strubell-etal-2019-energy,patterson2021carbon, henderson2020towards,anthony2020carbontracker,codecarbon2024, oviedo2025energyuseaiinference}. Most directly related, \citet{fernandez2025energy} jointly benchmark accuracy and energy on cloud GPUs (A100/H100) for fixed-format NLP tasks (sentiment classification, extractive QA, NLI). Our work is complementary but disjoint in scope: we study \textit{local} accelerators (M4 Max, RTX 6000, MI300X, smartphone NPUs) and naturalistic real-world query distributions, decompose efficiency longitudinally into model versus hardware contributions across successive generations, and analyze hybrid local-cloud routing, none of which are addressed by their cloud-only, fixed-task setup. Complementary to our focus on local-cloud routing, cost and efficiency-driven model selection strategies such as FrugalML and FrugalGPT for API and model cascades, and CALM for token-wise early exit, dynamically allocate workloads to cheaper or smaller models while preserving quality \citep{chen2020frugalml,chen2023frugalgpt,schuster2022calm}. On-device and edge studies demonstrate algorithm-hardware co-design for lower latency and energy consumption, exemplified by EdgeBERT's optimizations and PowerInfer's efficient LLM serving on commodity GPUs \citep{tambe2021edgebert,song2024powerinfer}. Finally, hardware-aware benchmarking efforts such as ``From Words to Watts'' and MLPerf Power quantify inference energy across accelerators and standardize power measurement protocols \citep{samsi2023words,tschand2025mlperfpower}.

\section{Dataset and Profiling Harness}
\label{app:dataset}
In this section, we provide additional details on our dataset curation for our study of hybrid local-cloud LM systems.

\subsection{Dataset Curation}
Here, we provide additional details on the Anthropic Economic Index~\citep{handa2025economictasksperformedai} categories used as labels (see Table~\ref{tab:anthropic_economic_index_categories}), the hardware platforms profiled, and the metrics recorded in our curated dataset.

\begin{table}[ht]
\centering
\small
\setlength{\tabcolsep}{6pt}
\renewcommand{\arraystretch}{1.2}
\begin{tabular}{@{}p{0.47\linewidth} p{0.47\linewidth}@{}}
\hline
Life, physical, and social science & Computer and mathematical \\
Architecture and engineering & Education instruction and library \\
Installation, maintenance, and repair & Business and financial operations \\
Legal services & Transportation and material moving \\
Arts, design, sports, entertainment, and media & Production services \\
Farming, fishing, and forestry & Healthcare support \\
Food preparation and serving related & Healthcare practitioners and technical \\
Community and social service & Sales and related \\
Office and administrative support & General management \\
Protective service & Building grounds cleaning and maintenance \\
Construction and extraction & Personal care and service \\
\hline
\end{tabular}
\caption{\textbf{Anthropic Economic Index Categories \citep{handa2025economictasksperformedai}}. This taxonomy categorizes occupations into 22 standardized economic domains, adapted from U.S. Bureau of Labor Statistics frameworks. It is designed to support AI impact analysis by aligning labor categories with distinct task structures.}
\label{tab:anthropic_economic_index_categories}
\end{table}

\paragraph{Query Curation}
\label{app:query_source_cleaning}
When sourcing queries from the \wildchat and \nr datasets, we apply robust data cleaning and filtering to ensure the quality and consistency of the sampled queries. For \nr, we filter out all queries that don't contain ground truth answers. For \wildchat, we eliminate non-English entries to maintain linguistic uniformity across the dataset. Queries that are malformed, nonsensical, or otherwise unintelligible (as determined by an LLM judge, i.e., \textsc{GPT-4o-mini}) are discarded to prevent noise. Additionally, duplicate queries are removed to reduce redundancy and avoid overrepresentation of specific prompts. Finally, we filter out excessively long queries that exceed a 32,000-character limit.

\paragraph{Dataset Statistics}

\autoref{tab:trafficbench_complete_stats_20b_appendix} reveals significant differences in how the two datasets are distributed across domains.
\wildchat is dominated by ``Arts, design, sports, entertainment, and media'' queries (47.1\%), followed by ``Computer and mathematical'' (18.1\%), while \nr is primarily composed of ``Life, physical, and social science'' (36.0\%) and ``Computer and mathematical'' (34.8\%) queries.
We use \textsc{gpt-4o-mini} to bucket each query into its economic categorization using the prompt below.

\begin{lstlisting}[basicstyle=\small\ttfamily,
  breaklines=true,
  backgroundcolor=\color{gray!10},
  frame=single]
You are a query categorizer. Your task is to categorize the following user query into one of the predefined categories based on the job/occupation domain it relates to most closely.

  Query: "{query}"

  Available Categories:
  - Office and administrative support
  - Transportation and material moving
  - Sales and related
  - Food preparation and serving related
  - General management
  - Business and financial operations
  - Healthcare practitioners and technical
  - Production services
  - Education instruction and library
  - Healthcare support
  - Construction and extraction
  - Installation, maintenance, and repair
  - Computer and mathematical
  - Building grounds cleaning and maintenance
  - Protective service
  - Personal care and service
  - Architecture and engineering
  - Community and social service
  - Arts, design, sports, entertainment, and media
  - Life, physical, and social science
  - Legal services
  - Farming, fishing, and forestry
  - None

  Instructions:
  1. Read the query carefully
  2. Determine which job/occupation category the query relates to most closely
  3. If the query doesn't clearly relate to any specific occupation category, use "None"
  4. Respond with ONLY the category name, exactly as listed above

  Category:
\end{lstlisting}

Solvability rates vary dramatically by domain and dataset type, where a query's solvability is defined as its ability to be answered correctly by any of the available local LMs (e.g. Qwen models or GPT OSS).
\wildchat queries show consistently high solvability across most domains (generally $>94\%$), with particularly strong performance in creative and social domains.
In contrast, \nr exhibits more variable solvability, with technical domains like ``Architecture and engineering'' showing only 41.5\% solvability compared to 99.4\% for the same domain in \wildchat.
This disparity reflects the complexity difference between open-ended chat queries and analytical reasoning tasks, supporting our findings that chat queries are more amenable to local model routing than reasoning-intensive queries.

\begin{figure}[htbp]
    \centering
    \includegraphics[width=0.9\textwidth]{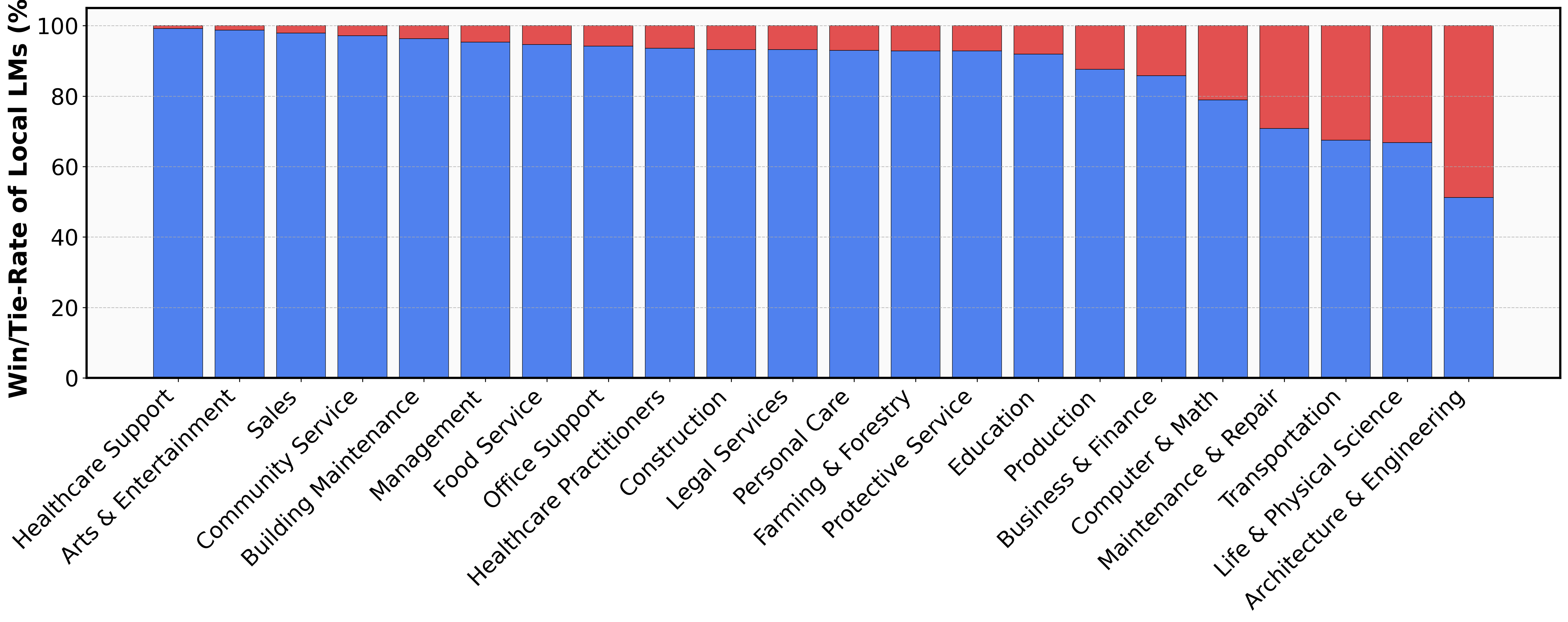}
    \caption{\textbf{Local Win/Tie-Rate vs. Cloud LMs by Domain}.  Stacked bars show the fraction of single-turn chat and reasoning queries handled by local LMs ($<20B$ active parameters; blue) versus those routed to frontier models in the cloud (red), computed per economic index domain~\citep{appelmccrorytamkin2025geoapi}}
    \label{fig:routing_share_by_domain}
\end{figure}

\begin{table}
  \centering
  \scriptsize
  \begin{tabular}{l@{\hspace{4pt}}r@{\hspace{4pt}}r@{\hspace{4pt}}r@{\hspace{4pt}}r@{\hspace{4pt}}r@{\hspace{4pt}}r}
  \hline
  \textbf{Domain} & \makecell{\textbf{WC} \\ \textbf{Count}} & \makecell{\textbf{WC} \\ \textbf{\%}} & \makecell{\textbf{WC} \\ \textbf{Solv \%}} & \makecell{\textbf{NR} \\ \textbf{Count}} & \makecell{\textbf{NR} \\ \textbf{\%}} & \makecell{\textbf{NR} \\ \textbf{Solv \%}} \\
  \hline
  Computer and mathematical & 90,662 & 18.1 & 99.5 & 174,242 & 34.8 & 67.3 \\
  \makecell[l]{Arts, design, sports, entertainment,\\ and media} & 235,658 & 47.1 & 98.7 & 2,648 & 0.5 & 52.9 \\
  \makecell[l]{Life, physical, and\\ social science} & 28,079 & 5.6 & 98.8 & 180,065 & 36.0 & 60.5 \\
  None & 49,014 & 9.8 & 97.3 & 79,752 & 16.0 & 65.6 \\
  Education instruction and library & 23,196 & 4.6 & 97.2 & 13,864 & 2.8 & 80.4 \\
  Architecture and engineering & 5,782 & 1.2 & 98.9 & 28,762 & 5.8 & 40.8 \\
  Business and financial operations & 19,628 & 3.9 & 97.8 & 8,779 & 1.8 & 55.3 \\
  \makecell[l]{Healthcare practitioners\\ and technical} & 8,905 & 1.8 & 98.1 & 1,851 & 0.4 & 66.3 \\
  \makecell[l]{Office and administrative\\ support} & 6,959 & 1.4 & 91.6 & 25 & 0.0 & 48.0 \\
  Legal services & 5,208 & 1.0 & 98.6 & 1,349 & 0.3 & 69.0 \\
  Community and social service & 6,125 & 1.2 & 97.0 & 404 & 0.1 & 76.2 \\
  \makecell[l]{Transportation and\\ material moving} & 1,890 & 0.4 & 95.1 & 3,914 & 0.8 & 53.0 \\
  Sales and related & 4,689 & 0.9 & 97.8 & 218 & 0.0 & 67.4 \\
  \makecell[l]{Food preparation and\\ serving related} & 3,308 & 0.7 & 98.3 & 507 & 0.1 & 61.9 \\
  General management & 3,364 & 0.7 & 96.7 & 340 & 0.1 & 67.9 \\
  \makecell[l]{Installation, maintenance,\\ and repair} & 954 & 0.2 & 97.1 & 2,037 & 0.4 & 56.2 \\
  Farming, fishing, and forestry & 1,677 & 0.3 & 99.5 & 411 & 0.1 & 65.5 \\
  Protective service & 1,315 & 0.3 & 97.9 & 237 & 0.0 & 56.5 \\
  Construction and extraction & 991 & 0.2 & 97.2 & 147 & 0.0 & 60.5 \\
  Healthcare support & 1,112 & 0.2 & 97.5 & 12 & 0.0 & 100.0 \\
  Production services & 546 & 0.1 & 100.0 & 334 & 0.1 & 65.3 \\
  Personal care and service & 648 & 0.1 & 92.9 & 19 & 0.0 & 0.0 \\
  \makecell[l]{Building grounds cleaning\\ and maintenance} & 278 & 0.1 & 100.0 & 70 & 0.0 & 72.9 \\
  \hline
  \textbf{TOTAL} & \textbf{500K} & \textbf{100.0} & \textbf{98.4} & \textbf{500K} & \textbf{100.0} & \textbf{63.0} \\
  \hline
  \end{tabular}
  \caption{\textbf{Dataset Domain Composition and LM Coverage ($\leq20B$ Active Parameter Models)}. Comparison of domain distribution and model solvability rates across \wildchat (WC) and \nr (NR) datasets. Solvability indicates the percentage of problems that can be solved correctly by at least one model with $\leq20B$ active parameters.}
  \label{tab:trafficbench_complete_stats_20b_appendix}
  \end{table}

\begin{table}[h]
\centering
\scriptsize
\begin{tabular}{p{6cm}cccc}
\toprule
\textbf{Category} & \textbf{\wildchat{}} & \textbf{\mmlupro{}} & \textbf{\supergpqa{}} & \textbf{Average} \\
\midrule
Computer and mathematical & 93.4\% & 90.6\% & 72.8\% & 85.6\% \\
Life, physical, and social science & 91.1\% & 84.7\% & 50.4\% & 75.4\% \\
Sales and related & 86.8\% & 74.2\% & 64.3\% & 75.1\% \\
Business and financial operations & 89.3\% & 82.9\% & 52.5\% & 74.9\% \\
Production services & 89.7\% & 85.7\% & 48.8\% & 74.7\% \\
Office and administrative support & 88.5\% & 83.3\% & 44.8\% & 72.2\% \\
Healthcare practitioners and technical & 88.9\% & 78.8\% & 48.0\% & 71.9\% \\
Installation, maintenance, and repair & 90.4\% & 80.0\% & 44.4\% & 71.6\% \\
Architecture and engineering & 90.0\% & 73.2\% & 51.6\% & 71.6\% \\
Protective service & 88.6\% & 75.0\% & 47.6\% & 70.4\% \\
Education instruction and library & 90.6\% & 77.4\% & 43.2\% & 70.4\% \\
Farming, fishing, and forestry & 87.2\% & 77.8\% & 45.9\% & 70.3\% \\
None & 86.5\% & 77.4\% & 42.4\% & 68.8\% \\
General management & 88.2\% & 76.3\% & 41.6\% & 68.7\% \\
Transportation and material moving & 91.1\% & 66.7\% & 44.9\% & 67.5\% \\
Construction and extraction & 91.0\% & 66.7\% & 41.4\% & 66.4\% \\
Food preparation and serving related & 86.1\% & 82.6\% & 30.3\% & 66.3\% \\
Community and social service & 87.4\% & 64.1\% & 45.5\% & 65.7\% \\
Arts, design, sports, entertainment, and media & 84.8\% & 75.5\% & 36.2\% & 65.5\% \\
Building grounds cleaning and maintenance & 90.1\% & 71.4\% & 30.8\% & 64.1\% \\
Legal services & 87.1\% & 61.5\% & 43.6\% & 64.1\% \\
Healthcare support & 86.5\% & 60.0\% & 38.9\% & 61.8\% \\
Personal care and service & 89.4\% & 50.0\% & 25.0\% & 54.8\% \\
\bottomrule
\end{tabular}
\caption{\textbf{GPT-OSS-120B Performance across Datasets.} Performance metrics across \wildchat{}, \nr{}, \mmlupro{}, and \supergpqa{} benchmarks, organized by Anthropic Economic Index categories \citep{handa2025economictasksperformedai}.}
\label{tab:gpt_120B_performance_across_datasets}
\end{table}

\paragraph{Metrics}
\label{app:metrics}
We detail all the metrics collected via our profiling harness in Table~\ref{tab:full_metrics_profile}.  For our correctness evaluations on \wildchat, we use an LLM-as-a-judge approach to evaluate model generated answers against a ground truth answer from \qwentwothirtyfive.
For our correctness evaluations of \nr, we use another LLM-judge prompt, but compare against ground truth answers provided in the original dataset. We provide both LLM-judge prompts below. The LLM used for each respective evaluation is \textsc{GPT-4o}.
For \supergpqa{} and \mmlupro{}, we simply compare the multiple choice answer selected in the response to the multiple choice answer of the reference response.

\textbf{\wildchat LLM-judge Prompt}
\begin{lstlisting}[basicstyle=\small\ttfamily,
  breaklines=true,
  backgroundcolor=\color{gray!10},
  frame=single]

You are an impartial judge evaluating the quality of two AI-assistant replies to the same user prompt.

Step 1: Generate your own answer
Write the response *you* would give to the user. Keep it separate from later analysis.

Step 2: Decide the query type
Classify the user prompt as either
- **Subjective / open-ended** (creative writing, opinion, advice, brainstorming)
- **Objective / technical** (code, math, logical derivations with a single correct outcome)
If uncertain, default to "Subjective".

Step 3 - Score each assistant with the correct rubric

| Query type | Criteria |
|------------|----------|
| Subjective / open-ended | 1. Correctness / factual soundness 2. Helpfulness 3. Relevance 4. Conciseness 5. Creativity & novelty |
| Objective / technical   | 1. Correctness only |

When using the multi-criteria rubric, note strengths and weaknesses for **each** dimension.
When using the single-criterion rubric, focus exclusively on factual / functional accuracy and ignore style or flair.

Step 4: Compare & justify
Explain which assistant is better and why, correcting any mistakes you find. Highlight missing but important details. **Be concise.**

Step 5:  Verdict
1. Assistant A is significantly better: [[A>>B]]
2. Assistant A is slightly better: [[A>B]]
3. Tie, Assistant A is equal: [[A=B]]
4. Assistant B is slightly better: [[B>A]]
5. Assistant B is significantly better: [[B>>A]]

Choose exactly one token from: `[[A>>B]]`, `[[A>B]]`, `[[A=B]]`, `[[B>A]]`, `[[B>>A]]`.

---

### Output format (strict)
Return **only** a JSON object that matches the provided schema:
\end{lstlisting}

\textbf{\nr LLM-judge Prompt}
\begin{lstlisting}[basicstyle=\small\ttfamily,
  breaklines=true,
  backgroundcolor=\color{gray!10},
  frame=single]
You are evaluating a response to a scientific/technical question against a reference answer.

Your task is to determine if the response is factually correct and complete compared to the reference.

Consider:
1. Scientific accuracy of facts and concepts
2. Mathematical correctness (if applicable)
3. Completeness of the answer
4. Technical precision

Question: {question}

Response: {response}

Reference Answer: {reference}

Return ONLY 'True' if the response is correct and complete, 'False' otherwise.
\end{lstlisting}

Table~\ref{tab:full_metrics_profile} catalogs the per-query metrics collected by our profiling harness across compute, energy, latency, memory, and utilization dimensions; together these constitute the raw measurements from which all IPW, IPJ, PPW, and PPJ figures in the main paper are derived.

\begin{table}[ht]
\centering
\begin{tabular}{|l|p{5.5cm}|}
\hline
\textbf{Metric} & \textbf{Description} \\
\hline
\texttt{flops\_per\_request} & FLOPs per query. \\
\texttt{macs\_per\_request} & MACs per query; proxy for compute. \\
\texttt{per\_query\_joules} & Energy per query (J). \\
\texttt{total\_joules} & Total energy across queries. \\
\texttt{per\_token\_ms} & Latency per token (ms). \\
\texttt{throughput\_tokens\_per\_sec} & Token output rate (toks/s). \\
\texttt{time\_to\_first\_token\_seconds} & Time to first token (s). \\
\texttt{total\_query\_seconds} & Total time per query (s). \\
\texttt{cpu\_mb.avg / max / median / min} & CPU memory usage (MB). \\
\texttt{gpu\_mb.avg / max / median / min} & GPU memory usage (MB). \\
\texttt{initialization\_duration\_seconds} & Model load time (s). \\
\texttt{batch\_size} & Query batch size. \\
\texttt{gpu\_memory\_utilization} & GPU memory use (0--1). \\
\texttt{max\_model\_len} & Max token length allowed. \\
\texttt{max\_num\_batched\_tokens} & Max batch token count. \\
\texttt{max\_output\_tokens} & Max output tokens. \\
\texttt{num\_workers} & Number of threads. \\
\texttt{temperature} & Sampling temperature. \\
\texttt{top\_k} & Top-k cutoff. \\
\texttt{top\_p} & Top-p (nucleus) threshold. \\
\texttt{warmup\_steps} & Warm-up steps. \\
\texttt{per\_query\_watts.avg / max / median / min} & GPU power draw per query (W). \\
\texttt{total\_watts.avg / max / median / min} & Session GPU power draw (W). \\
\texttt{cpu\_count} & CPU core count. \\
\texttt{cpu\_brand} & CPU model. \\
\texttt{host\_name} & Machine hostname. \\
\texttt{os\_name / os\_version / kernel\_version} & OS and kernel info. \\
\texttt{temperature.avg / max / median / min} & Device temperature ($^\circ$C). \\
\texttt{input} & Input tokens per query. \\
\texttt{output} & Output tokens per query. \\
\hline
\end{tabular}
\caption{\textbf{Dataset Metrics}. Summary of compute, latency, memory, and energy profiling metrics.}
\label{tab:full_metrics_profile}
\end{table}

\paragraph{Hardware Backends}
Details regarding profiled hardware can be found in Table~\ref{tab:accelerator_details}.

\paragraph{Data Generation Procedure}
\label{app:data_generation_procedure}
We generate model outputs using consistent decoding settings across all tasks: \texttt{temperature $=$ 0.6}, \texttt{top-p $=$ 0.95}, \texttt{top-k $=$ 20}, \texttt{min-p $=$ 0.0}, and a 32768-token output limit. For \nr, \supergpqa and \gpqa queries, we enable deliberative prompting (\texttt{use thinking $=$ True}); for \wildchat, we disable it. For \textsc{Qwen} models, we apply a repetition penalty of 1.1 and length penalty of 1.0.

\begin{table}[h]
\centering
\scriptsize
\setlength{\tabcolsep}{4pt}
\begin{tabular}{l l l l}
\hline
\textbf{Hardware} & \textbf{Memory} & \textbf{Bandwidth} & \textbf{Power} \\
\hline
NVIDIA A100 (Ampere)            & 40 GB HBM2     & 1{,}555 GB/s   & 400 W TDP \\
NVIDIA H200 (Hopper)             & 141 GB HBM3e   & 4.8 TB/s       & Up to 700 W TDP \\
NVIDIA B200 (Blackwell)          & 192 GB HBM3e   & 8 TB/s         & 1000 W TDP \\
NVIDIA GH200 (Grace Hopper)      & 144 GB HBM3e (+624 GB LPDDR5X) & 4.8 TB/s (GPU) & 1000 W TDP \\
NVIDIA Quadro RTX 6000 (Turing)  & 24 GB GDDR6    & 672 GB/s       & 295 W TDP \\
NVIDIA RTX 6000 Ada Generation   & 48 GB GDDR6    & 960 GB/s       & 300 W TDP \\
AMD Instinct MI300X (CDNA 3)     & 192 GB HBM3    & 5.3 TB/s (peak) & 750 W TBP \\
Apple Mac Studio (M4 Max)        & 128 GB unified & 546 GB/s       & 480 W (system PSU)\textsuperscript{$\dagger$} \\
Apple iPhone 16 Pro (A18 Pro)    & 8 GB LPDDR5X   & 60 GB/s        & ${\sim}12$ W (SoC peak) \\
SambaNova SN40L RDU              & 64 GB HBM2E    & 1.6 TB/s       & 500 W TDP \\
\hline
\end{tabular}
\caption{\textbf{Accelerator Details}. Memory, bandwidth, and power specifications of evaluated accelerators and systems. \textsuperscript{$\dagger$}For the Apple Mac Studio (M4 Max), 480~W is the system power-supply continuous rating; the M4 Max SoC under sustained AI workloads draws substantially less. Per-query power measurements in our study are taken via \texttt{powermetrics} \texttt{processor\_power.actual} (GPU subsystem only) so that comparisons against NVML/ROCm-SMI accelerator-only power on other platforms are like-for-like (see App.~\ref{app:telemetry_collection}).}
\label{tab:accelerator_details}
\end{table}

\paragraph{Telemetry Collection}
\label{app:telemetry_collection}
We collected telemetry by instrumenting host-level samplers that interface directly with vendor-supported system APIs on each platform. Data were obtained from NVML on NVIDIA-equipped hosts, from the \texttt{powermetrics} facility on macOS, and from ROCm SMI on AMD-equipped hosts. Each sampler queried the respective system interface to obtain GPU- and system-level measurements and produced synchronized records suitable for downstream quantitative analysis.

On NVIDIA systems, we interface directly with NVML and enumerate all visible GPUs. For each device, we query instantaneous power as reported by the driver, read cumulative energy from the on-device counter, obtain GPU temperature from the hardware sensor, and retrieve memory usage from the device's memory interface. Units are normalized (e.g., milliwatts and millijoules mapped to watts and joules; bytes to megabytes). In multi-GPU hosts, power and memory are summed across devices and temperature is averaged to yield a single aggregate view. Each record also includes host memory usage from OS counters, a nanosecond timestamp, and device identity and backend provenance.

On macOS, we execute \texttt{powermetrics} with elevated privileges and ingest its continuous \texttt{plist} stream. Each \texttt{plist} frame is parsed to extract the GPU power value exposed by the system (Apple Silicon: \texttt{processor\_power.actual}; Intel: \texttt{processor.combined\_power}), which is normalized to watts. Energy (joules) is obtained by numerically integrating the power signal over successive frames using the measured inter-frame wall-clock interval. In parallel, system memory usage is sampled from OS counters. Every observation is timestamped and annotated with Apple device identity and an explicit \texttt{powermetrics} backend tag.

On AMD systems, we use ROCm SMI to query current GPU power (watts), read temperature from junction or edge sensors ($^\circ$C), and obtain VRAM usage from the device memory interface (bytes to megabytes). Energy (joules) is computed by integrating the power signal over time using consecutive sampling intervals. System memory usage is read from OS counters. In multi-GPU machines, the primary device under observation is explicitly selected (GPU index 0 in our setup), and all records carry precise timestamps together with device identity and backend metadata.

To ensure measurement precision and account for variance in inference-time behavior, we execute each query 10 times and aggregate power measurements across runs. For each query, we compute the mean power draw (watts) and mean energy consumption (joules) per query by averaging across these 10 independent executions. This repeated sampling approach reduces measurement noise and provides robust estimates of per-query resource consumption that account for system-level variability in accelerator utilization, thermal conditions, and memory allocation patterns.

\begin{table}[t]
\centering
\scriptsize
\begin{tabular}{l|cc|cc|cc}
\toprule
 & \multicolumn{2}{c|}{\textbf{Cost Savings}} & \multicolumn{2}{c|}{\textbf{Compute Savings}} & \multicolumn{2}{c}{\textbf{Energy Savings}} \\
\cmidrule{2-3} \cmidrule{4-5} \cmidrule{6-7}
\textbf{Size Threshold ($\leq$)} & \makecell{Qwen \\ + GPT-OSS} & {Qwen} & \makecell{Qwen \\ + GPT-OSS} & {Qwen} & \makecell{Qwen \\ + GPT-OSS} & {Qwen} \\
\midrule
4B   & 65.2\% & 65.2\% & 65.1\% & 65.1\% & 63.5\% & 63.5\% \\
8B   & 80.8\% & 80.8\% & 83.1\% & 83.1\% & 79.6\% & 79.6\% \\
14B  & 89.0\% & 89.0\% & 93.0\% & 93.0\% & 87.0\% & 87.0\% \\
20B  & 90.5\% & ---    & 97.4\% & ---    & 89.4\% & ---    \\
32B  & 91.3\% & 91.9\% & 97.4\% & 92.8\% & 90.4\% & 90.5\% \\
120B & 91.1\% & ---    & 97.7\% & ---    & 90.7\% & ---    \\
\bottomrule
\end{tabular}
\caption{\textbf{Cost, Compute, and Energy Savings from Local-Cloud Routing on WildChat}: Savings across different resources while maintaining task accuracy of SOTA open-source cloud model (i.e. \texttt{Qwen3 235B-A22B}).}
\label{tab:wildchat_savings}
\end{table}

\begin{table}[t]
\centering
\scriptsize
\begin{tabular}{l|cc|cc|cc}
\toprule
 & \multicolumn{2}{c|}{\textbf{Cost Savings}} & \multicolumn{2}{c|}{\textbf{Compute Savings}} & \multicolumn{2}{c}{\textbf{Energy Savings}} \\
\cmidrule{2-3} \cmidrule{4-5} \cmidrule{6-7}
\textbf{Size Threshold ($\leq$)} & \makecell{Qwen \\ + GPT-OSS} & {Qwen} & \makecell{Qwen \\ + GPT-OSS} & {Qwen} & \makecell{Qwen \\ + GPT-OSS} & {Qwen} \\
\midrule
4B   & 52.9\% & 52.9\% & 54.5\%    & 54.5\%    & 46.3\% & 46.3\% \\
8B   & 60.5\% & 60.5\% & 62.5\%    & 62.5\%    & 54.0\% & 54.0\% \\
14B  & 68.7\% & 68.7\% & 70.1\%    & 70.1\%    & 62.5\% & 62.5\% \\
20B  & 73.3\% & ---    & 72.2\%    & ---    & 67.8\% & ---    \\
32B  & 76.9\% & 75.9\% & 75.1\%    & 75.1\%    & 72.4\% & 71.6\% \\
120B & 86.7\% & ---    & 86.0\%    & 86.0\%    & 85.9\% & ---    \\
\bottomrule
\end{tabular}
\caption{\textbf{Cost, Compute, and Energy Savings from Local-Cloud Routing on NaturalReasoning}: Savings across different resources while maintaining task accuracy of SOTA open-source cloud model (i.e. \texttt{Qwen3 235B-A22B}).}
\label{tab:naturalreasoning_savings}
\end{table}

\section{Limitations and Broader Impacts (Extended)}
\label{app:limitations_and_broader_impacts}

\paragraph{Limitations.} Our study is subject to several limitations. \textbf{(1) Measurement precision.} Our energy and power measurements rely on software-level telemetry (NVML, \texttt{powermetrics}, ROCm SMI), which can introduce inaccuracies of $10$--$15\%$~\citep{yang2023part} and may not capture milliwatt-level variations that hardware wattage meters would. While our methodology aligns with established practices and provides consistent relative comparisons, absolute energy values should be interpreted accordingly. \textbf{(2) Query coverage.} Our analysis focuses on single-turn chat and reasoning queries. Multi-turn conversations, agentic workflows, tool use, and long-context applications represent substantial portions of real-world LLM traffic; we report a multi-turn extension on GAIA and TerminalBenchV2 in App.~\ref{app:multi_turn} that confirms the qualitative patterns generalize, but routing decisions and efficiency tradeoffs may still differ in those settings. \textbf{(3) Evaluation methodology.} Our correctness measurements rely on LLM-as-a-judge (\textsc{Qwen3-235B}) for open-ended chat queries, which inherits any biases or systematic errors of the judge model. \textbf{(4) Hardware coverage.} While we evaluate eight datacenter, workstation, and consumer-class accelerators (and additionally a smartphone-class accelerator in App.~\ref{app:smartphone_experiments}), we do not cover the full diversity of available local hardware (e.g., other mobile NPUs, integrated GPUs, edge accelerators), and our findings on local efficiency may not extrapolate uniformly across these classes. \textbf{(5) Batch size = 1 cloud comparison.} Our main per-query comparisons (Tables~\ref{tab:apple_m4_max_vs_nvidia_b200_for_power}--\ref{tab:apple_m4_max_vs_nvidia_b200_for_energy}) use batch size 1 to follow standard local-inference benchmarking practice~\citep{hao2023reaching}. This is conservative for cloud accelerators: App.~\ref{app:batching_ablation} shows bs=64 on B200 yields $11$--$20\times$ higher IPJ. The routing simulation (Section~\ref{sec:efficiency_gains_from_routing}) uses bs=16 for the cloud baseline. \textbf{(6) Single-runtime design.} We standardize on \textsc{vLLM} to isolate model-hardware effects; App.~\ref{app:framework_sensitivity} shows absolute IPW shifts by $3$--$12\%$ across \textsc{vLLM}/\textsc{SGLang}/\textsc{llama.cpp} but rankings are preserved (Kendall's $\tau \in [0.87, 0.93]$). System-level work such as Zeus~\citep{you2023zeus} and uServe~\citep{qiu2024userve} explores complementary axes (scheduling, power management) that we do not vary. \textbf{(7) Ecosystem velocity.} Our findings reflect models and accelerators available as of October 2025; sustained increases in DRAM/HBM pricing could slow the pace at which larger models become locally deployable, and our open-source profiling harness is designed to support periodic re-evaluation.

\paragraph{Broader Impacts.} Demonstrating that local inference can serve a substantial fraction of LLM queries has potential benefits for energy consumption, infrastructure cost, and access to AI capabilities, particularly in settings where cloud connectivity is unreliable, expensive, or undesirable for privacy reasons. Our findings are situated within a rapidly accelerating local-AI ecosystem. On the software side, a wave of open-source personal-AI agent stacks has emerged that treats on-device execution as a design constraint rather than a fallback. Projects span the full hardware spectrum: \textsc{OpenClaw}~\citep{steinberger2025openclaw} and \textsc{ZeroClaw}~\citep{zeroclaw2026} target workstation deployment; \textsc{NanoBot}~\citep{hkuds2026nanobot}, \textsc{TinyClaw}~\citep{gonzaga2026tinyclaw}, and \textsc{Hermes Agent}~\citep{nousresearch2025hermes} focus on lightweight runtimes with cross-session memory and skill systems; \textsc{IronClaw}~\citep{nearai2026ironclaw} emphasizes privacy and security; \textsc{PicoClaw}~\citep{sipeed2026picoclaw} runs in under 10~MB of RAM on \$10 RISC-V hardware; and \textsc{MimiClaw}~\citep{memovai2026mimiclaw} demonstrates a full agent on a \$5 ESP32 microcontroller. Recent academic and industrial efforts include \textsc{OpenJarvis}~\citep{saadfalcon2026openjarvis}, which integrates local model serving with agent and memory components. These projects illustrate growing interest in deployment patterns where local inference handles a substantial share of personal-AI queries. On the hardware side, purpose-built local AI hardware, e.g.\ NVIDIA DGX Spark (128~GB unified memory, 1 petaFLOP at FP4) and Dell Pro Max workstations with GB300 (748~GB, 20 petaFLOPS at FP4), is bringing datacenter-class AI to desktop devices for the first time. Hybrid local-cloud routing, as we show, can reduce inference energy by $60$--$80\%$ at platform scale; the IPW metric and profiling harness we release are intended as a shared evaluation framework for tracking efficiency across this growing ecosystem as it matures. Our profiling harness lowers the barrier for systematic energy benchmarking, which we view as a public good given the rising aggregate energy footprint of AI workloads. We also note potential negative implications. First, increased local deployment of capable LMs may make certain forms of misuse (e.g., generating misinformation or harassment) harder to monitor and mitigate compared to cloud-served inference, where providers can apply usage policies. Second, local inference redistributes energy consumption to consumer power grids rather than eliminating it, and results on intelligence per watt could be misused to justify deployment patterns (e.g., always-on local agents) that, in aggregate, raise total energy consumption despite per-query efficiency gains: a Jevons-paradox concern we have not quantified. Third, our cost and energy comparisons are approximate and could be cited out of context to support specific procurement or policy decisions for which finer-grained, deployment-specific analysis would be more appropriate.

\section{Future Work}
\label{app:future_work}

We outline three directions for extending this work.

\subsection{Broadening Workload Coverage}
\label{app:future_work_workloads}

\paragraph{Multi-turn and Agentic Workflows} Our main study targets single-turn queries, and App.~\ref{app:multi_turn} reports an extension on GAIA and TerminalBenchV2 confirming that local-cloud efficiency patterns generalize qualitatively. Future work should profile IPW across diverse agent stacks, longer trajectories, and workloads dominated by tool-call overhead. Routing in agentic settings must account for cumulative trajectory cost rather than per-turn cost, which may shift the local-cloud crossover relative to single-turn workloads.

\paragraph{Long-context Inference} Our evaluation caps inputs at $32{,}000$ characters, but production workloads increasingly involve $100$K--$1$M token contexts where prefill dominates total energy and KV-cache pressure becomes the binding constraint. Characterizing IPW as a function of context length would clarify how the prefill-decode energy split shifts with input size and how unified-memory and HBM architectures behave in such settings.

\paragraph{Multimodal Workloads} Our study is text-only. Vision-language, audio, and video inference have different compute and memory profiles: image and video tokenization shift the prefill-decode balance, and modality-specific encoders introduce energy costs not captured by text-only profiling. Extending our harness to multimodal models and characterizing local-hardware viability on these workloads is an important direction.

\subsection{Pushing the Local-AI Frontier}
\label{app:future_work_frontier}

\paragraph{Closing the Reasoning Gap} App.~\ref{app:tbench_difficulty_analysis} shows Level 5 reasoning queries remain $95\%$ unsolved by current local LMs, and the hardest reasoning problems show markedly slower year-over-year progress than chat or moderate reasoning. Future small-model architectures, reasoning-focused post-training, and test-time compute strategies that fit within local power budgets are needed to push the locally-serviceable share beyond $88.7\%$. Whether long chain-of-thought reasoning can be made energy-efficient enough for local deployment is an open question.

\paragraph{Specialized On-device Architectures} Today's local LMs are predominantly GPU-centric models repurposed for local hardware, and App.~\ref{app:local_vs_cloud_intelligence_efficiency} shows local accelerators trail cloud accelerators by $1.4$--$7.4\times$ in IPW on identical workloads. Closing this gap may require NPU-first model designs, sparsity-aware architectures, and co-designed model-hardware stacks. The smartphone-class results in App.~\ref{app:smartphone_experiments} (${\sim}7\times$ higher IPW than workstation GPUs at ${\sim}12$~W) suggest substantial headroom for purpose-built mobile deployments.

\paragraph{Quantization Frontiers Below FP4} App.~\ref{app:model_precision_vs_accuracy} shows FP16$\rightarrow$FP4 yields $3$--$3.5\times$ energy reduction at ${\sim}2.5$pp accuracy loss per step. Sub-FP4 regimes (INT2, ternary, binary) offer additional headroom but require both training recipes that preserve accuracy and dedicated hardware support that few accelerators currently provide. Characterizing the accuracy floor of extreme quantization and the hardware support needed to realize its theoretical gains are open directions.

\subsection{Operationalizing Hybrid Inference}
\label{app:future_work_hybrid}

\paragraph{Serving-stack and System-level Optimizations} We standardize on \textsc{vLLM} and bs=1 for local inference; App.~\ref{app:batching_ablation} and App.~\ref{app:framework_sensitivity} confirm batch size and framework (i.e., VLLM vs SGLANG) shift absolute IPW but preserve rankings. A full treatment of speculative decoding, paged attention, DVFS, power capping, and cross-application local batching could materially improve local IPW. Cross-application batching is particularly promising because single-user devices cannot batch across users but can aggregate concurrent queries across on-device applications.

\paragraph{Measuring IPW Under Hybrid Execution} Our IPW metric is defined per (model, accelerator) pair and characterizes local or cloud inference in isolation. Hybrid execution patterns---speculative decoding with a local draft and cloud verifier, edge-cloud model partitioning~\citep{li2025sled,xie2025hat}, and collaborative protocols like Minions~\citep{narayan2025minions}---split a single query across local and cloud infrastructure and require an extended IPW formulation that aggregates power and energy across heterogeneous accelerators.

\section{Local-Cloud Experiments}
\label{app:tbench_experiments}

\subsection{Key Takeaways for Practitioners}
\label{app:key_takeaways}

We consolidate our empirical findings into eight concrete decision rules, each grounded in a specific experiment.
\textbf{(1) Architecture.} On memory-rich local devices, MoE delivers the best IPW: \gpthundred{} ($\leq 20$B active) achieves the highest single-model coverage ($71.4\%$, \autoref{fig:query_coverage}) and best IPW (\autoref{tab:longitudinal_trends_for_local_cloud}); capacity-to-compute decoupling more than compensates for storing all experts.
\textbf{(2) Quantization.} FP16$\to$FP4 yields $3$--$3.5\times$ energy reduction at ${\sim}2.5$pp accuracy loss per step (App.~\ref{app:model_precision_vs_accuracy}); a larger model at FP4 typically beats a smaller model at FP16, so scale model size first and quantize aggressively.
\textbf{(3) Routing.} Invest in router accuracy up to ${\sim}80\%$ ($80\%$ of oracle gains); beyond that, expand the local-model ensemble: the $17.3$pp best-single vs.\ best-of-local gap (\autoref{fig:query_coverage}) shows diversity matters more than additional routing accuracy.
\textbf{(4) Domains.} Coverage exceeds $93\%$ for creative fields but drops to $60\%$ for Architecture \& Engineering (\autoref{fig:routing_share_by_domain}); model developers expanding local-AI viability should prioritize technical reasoning.
\textbf{(5) Hardware bottleneck.} Cloud accelerators with HBM3e and dedicated tensor units achieve $1.4$--$7.4\times$ higher efficiency than M4 Max's unified memory (Tables~\ref{tab:apple_m4_max_vs_nvidia_b200_for_power}--\ref{tab:apple_m4_max_vs_nvidia_b200_for_energy}); the dominant local bottleneck is memory bandwidth and specialized compute, not raw FLOPs.
\textbf{(6) Power envelope.} Smartphone-class NPUs occupy a distinct regime: the iPhone~16~Pro achieves $\sim$$7\times$ higher IPW than workstation GPUs at ${\sim}12$\,W (App.~\ref{app:smartphone_experiments}), motivating NPU-optimized mobile deployment for the lightest queries.
\textbf{(7) Serving stack.} Cloud bs=$64$ on B200 yields $11$--$20\times$ higher IPJ than bs=$1$ (App.~\ref{app:batching_ablation}); local single-user deployment cannot batch but can aggregate concurrent on-device applications.
\textbf{(8) Framework.} Absolute IPW shifts $3$--$12\%$ across vLLM, SGLang, llama.cpp but rankings are preserved (Kendall's $\tau \in [0.87, 0.93]$, App.~\ref{app:framework_sensitivity}); standardize for comparison, but benchmark on the target stack for absolute throughput.

\subsection{How has local LM task coverage changed over different ``difficulty'' slices of the data}
\label{app:tbench_difficulty_analysis}

Using labels for query difficulty we quantify the rate of improvement of local LMs across task difficulty slices. We label each query by the minimum model size (in parameters) required to solve it when considering the SOTA LMs as of August 2025, categorizing queries into five difficulty levels: level 1 ($\leq$ 4B params), level 2 ($\leq$ 8B params), level 3 ($\leq$ 20B params), level 4 ($\leq$ 235B params), and level 5 (unsolvable).

\begin{figure}[htbp]
    \centering
    \includegraphics[width=0.7\textwidth]{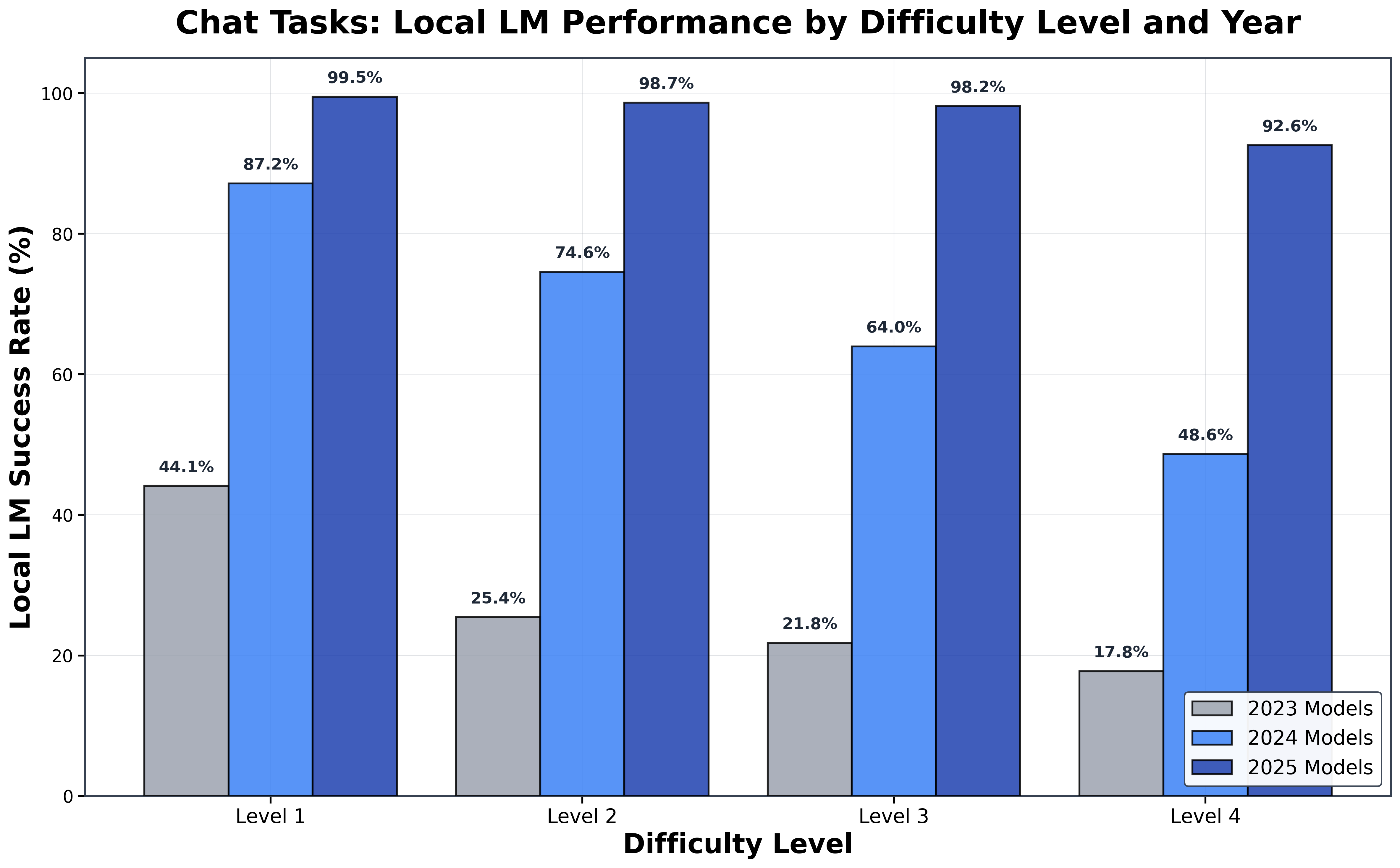}
        \caption{\textbf{Chat Task Performance by Difficulty Level and Year}. Model success rates across four difficulty levels and three model generations (2023, 2024, 2025). The data reveals dramatic progress across all difficulty levels, with 2023 models achieving 28.79\% overall success rising to 98.12\% by 2025. Notably, Levels 1-3 approach near-perfect performance (98-99\%), while Level 4 shows the largest relative improvement (+210.4\% per year) despite starting from the lowest baseline (17.77\%).}
    \label{fig:wildchat_difficulty_analysis}
\end{figure}

For \textbf{chat tasks} (see Figure~\ref{fig:wildchat_difficulty_analysis}), we observe near-universal performance gains across all difficulty levels, with 2025 models achieving 98-99\% success on levels 1-3 and 92.6\% on level 4. Absolute improvements range from +55.4 percentage points (pp) for level 1 to +76.4 pp for level 3, indicating relatively uniform capability gains.

\begin{figure}[htbp]
    \centering
    \includegraphics[width=0.7\textwidth]{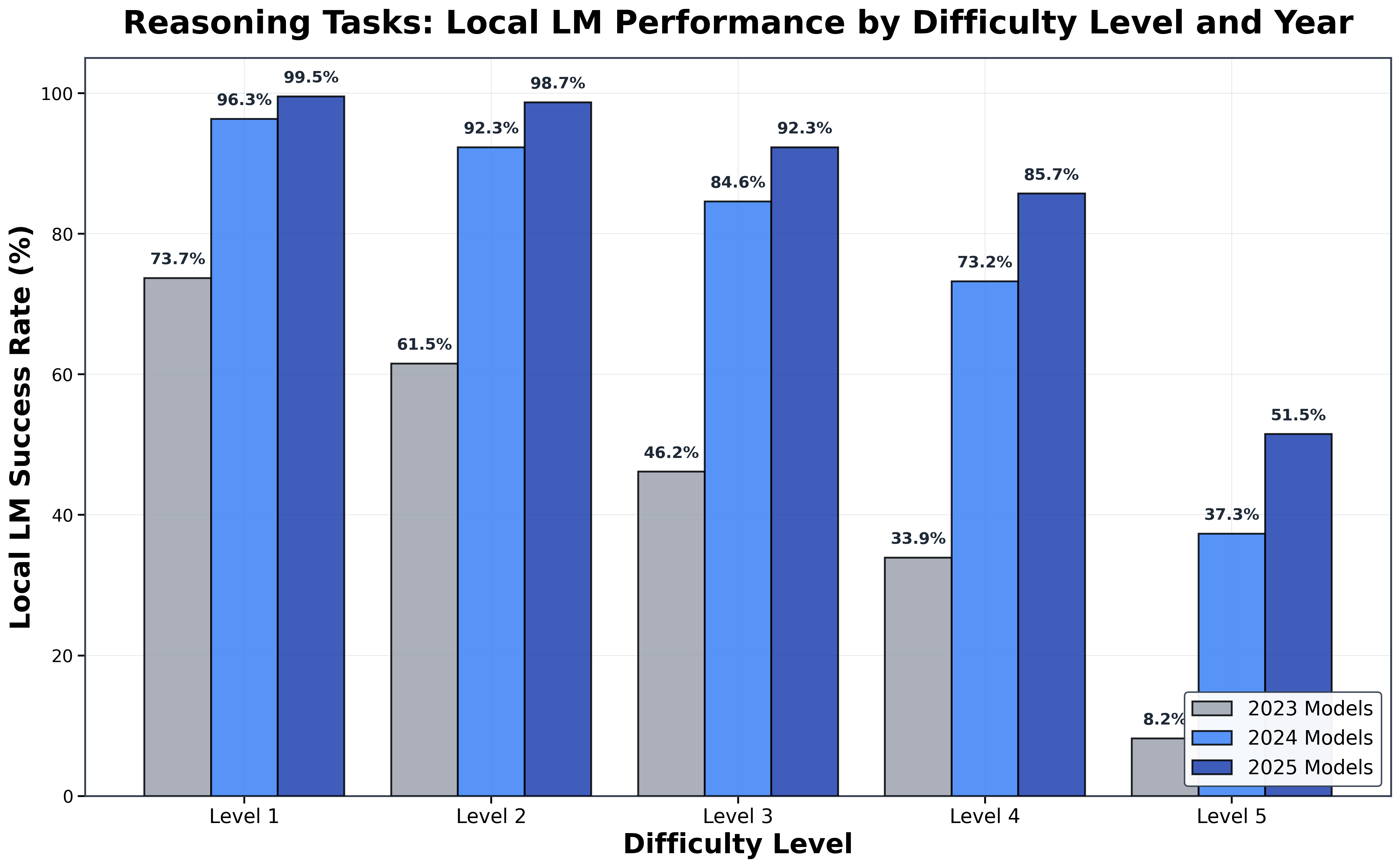}
        \caption{\textbf{Reasoning Task Performance by Difficulty Level and Year}. Model success rates on across five difficulty levels and three model generations. The benchmark shows a three-tier saturation pattern: near-complete (98-99\% on Levels 1-2), approaching saturation (85-92\% on Levels 3-4), and wide-open frontier (51\% on Level 5).}
    \label{fig:nr_difficulty_analysis}
\end{figure}

For \textbf{reasoning tasks} (see Figure~\ref{fig:nr_difficulty_analysis}, the pattern differs substantially. While levels 1-3 show strong improvements (+24.0, +37.8, and +53.9 pp respectively), levels 4 and 5 exhibit markedly slower progress. Level 4 improves by only +23.8 pp (7.93\% to 31.72\%), and level 5 remains largely unsolved with just +1.5 pp improvement (3.27\% to 4.72\%). This suggests that while local models have rapidly closed the gap on moderately difficult reasoning tasks, the hardest reasoning problems (those requiring either massive scale or capabilities beyond current architectures) remain a significant frontier. The presence of 134 level 5 problems (16.5\% of the reasoning dataset) that remain 95\% unsolved indicates substantial headroom for future model development in complex reasoning domains.

\FloatBarrier

\subsection{How much has intelligence efficiency changed over time?}
\label{app:additional_efficiency_trends}

\begin{figure*}[t]%
    \centering
    \includegraphics[width=1.0\textwidth]{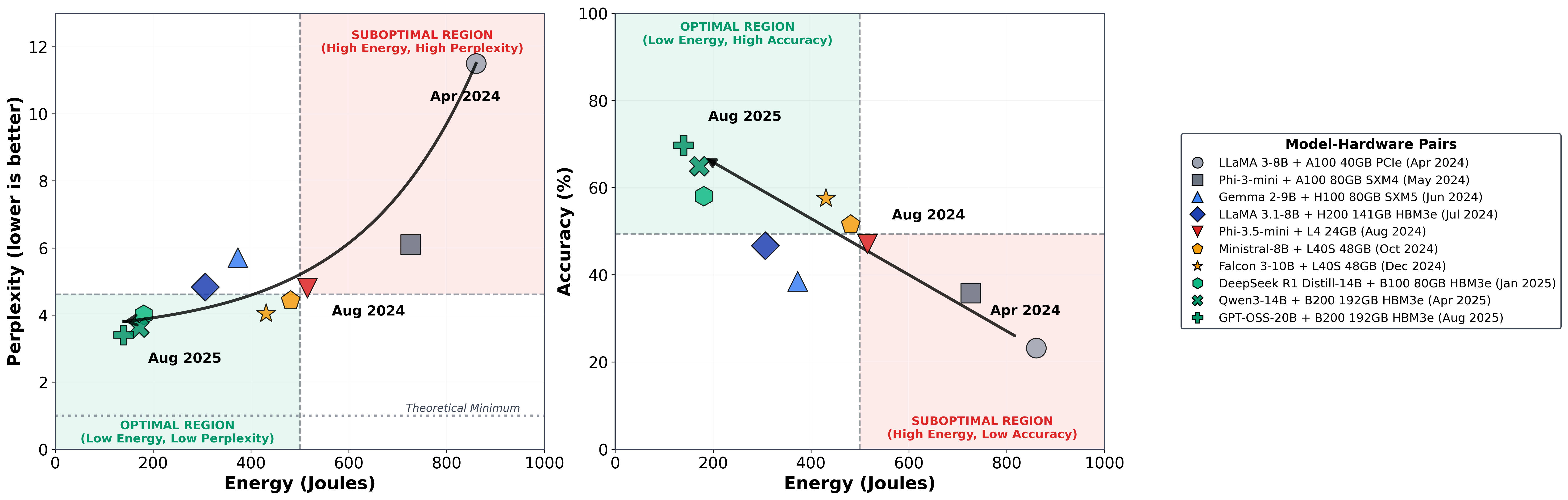}
    \caption{\textbf{Perplexity and Accuracy per Joule Trends across \wildchat{} and \nr{}.}}  %
    \label{fig:perplexity_and_accuracy_trends}
\end{figure*}

Figure~\ref{fig:perplexity_and_accuracy_trends} presents results across single-turn chat and reasoning queries, evaluating intelligence efficiency across model-hardware pairs from April 2024 through August 2025. We measure both perplexity (left panel) and accuracy (right panel) normalized by energy consumption in joules per query, tracking nine distinct model families (\textsc{Llama}, \textsc{Phi}, \textsc{Gemma}, \textsc{Mistral}, \textsc{Falcon}, \textsc{DeepSeek}, \textsc{Qwen}, and \textsc{GPT-OSS}) deployed on various GPU configurations including NVIDIA A100 (40GB/80GB PCIe/SXM), H100 (80GB SXM), H200 (141GB HBM3e), and L40S (48GB) accelerators. Energy measurements capture end-to-end inference costs. The temporal snapshots at April 2024, August 2024, and August 2025 enable direct comparison of efficiency trajectories, revealing how successive generations of models and hardware migrate from suboptimal regions (high energy, low performance) toward optimal regions (low energy, high performance) as indicated by the shaded quadrants in each panel.

\subsection{What efficiency gains can effective query routing deliver?}
\label{app:real_world_impact}
We compute cost per query using pricing available on OpenRouterAI~\citep{openrouter}. Table~\ref{tab:model_pricing_openrouter} lists the token pricing used.

\begin{table}[h]
\centering
\scriptsize
\begin{tabular}{lcc}
\toprule
\textbf{Model} & \textbf{Input Cost (USD / 1M tokens)} & \textbf{Output Cost (USD / 1M tokens)} \\
\midrule
Qwen3-4B & 0.000 & 0.000 \\
Qwen3-8B & 0.035 & 0.138 \\
Qwen3-14B & 0.060 & 0.124 \\
Qwen3-32B & 0.100 & 0.450 \\
Qwen3-235B & 0.220 & 0.880 \\
GPT-OSS-20B & 0.03 & 0.14 \\
GPT-OSS-120B & 0.15 & 0.60  \\
\bottomrule
\end{tabular}
\caption{\textbf{Model pricing from OpenRouterAI.} Costs are in USD per 1M tokens for input and output, as of August 2025.}
\label{tab:model_pricing_openrouter}
\end{table}

\subsection{How does model precision affect performance and efficiency?}
\label{app:model_precision_vs_accuracy}

Model quantization (reducing numerical precision from \textsc{Fp16} to \textsc{Fp8} or \textsc{Fp4}) decreases memory requirements and energy consumption during inference while introducing approximation error that may degrade model accuracy.
To quantify this tradeoff, we evaluate eight open-source models from the \textsc{Qwen3} and \textsc{Gemma} families across three precision levels: \textsc{Fp16} (full precision), \textsc{Fp8} (8-bit floating point), and \textsc{Fp4} (4-bit floating point). For \textsc{Fp8} and \textsc{Fp4} we use the vendor-published quantized checkpoints where available, falling back to native PyTorch FP8 / NVIDIA TransformerEngine FP4 conversions when no published checkpoint exists; energy measurements are taken on \textsc{NVIDIA B200} (cloud) and replicated on \textsc{Apple M4 Max} (local) using the same telemetry harness described in App.~\ref{app:telemetry_collection}, with per-query results averaged across hardware platforms.
For each model-precision pair, we measure accuracy on three reasoning-focused datasets: \nr{} ($N=10,000$), SuperGPQA ($N=10,000$), and MMLU Pro ($N=10,000$).

\begin{figure*}[t]%
    \centering
    \includegraphics[width=1.0\textwidth]{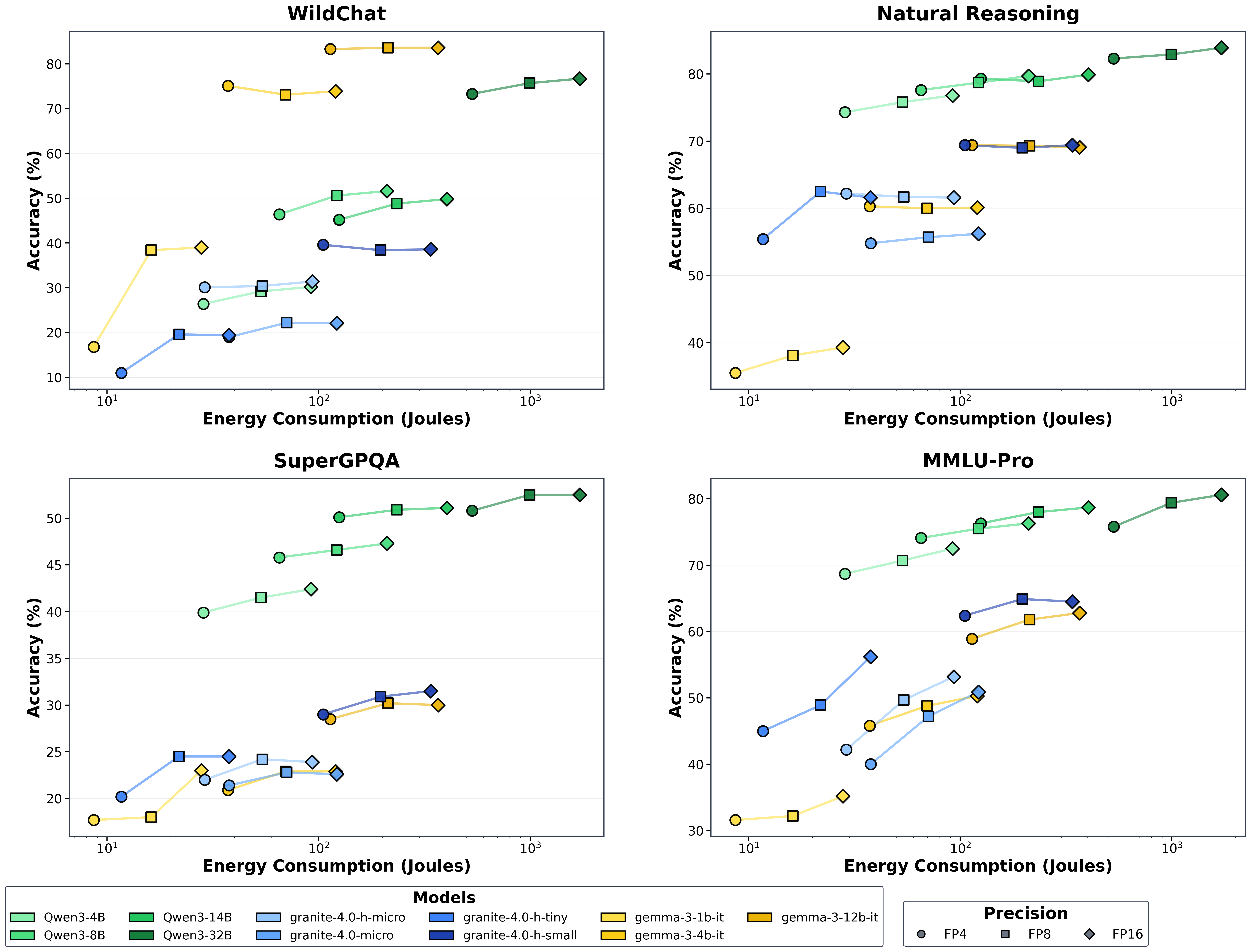}
    \caption{\textbf{Minimal Accuracy Degradation Shifting from FP16 to FP4 for Open-Source Local Models}: Evaluation across three reasoning datasets ($N=10,000$ each) shows $2-3\%$ accuracy loss per precision step, demonstrating that $FP8$/$FP4$ quantization enables efficient deployment with acceptable performance tradeoffs.}
    \label{fig:model_precision_vs_accuracy}
\end{figure*}

Figure~\ref{fig:model_precision_vs_accuracy} shows that quantization from \textsc{Fp16} to \textsc{Fp4} yields energy reductions of $3-3.5\times$ with accuracy degradation of approximately 2.5 percentage points per precision step across all models and datasets.
For example, on SuperGPQA, \textsc{Qwen3-14B} achieves 54.5\% (\textsc{Fp16}), 52.0\% (\textsc{Fp8}), and 49.0\% (\textsc{Fp4}): a total degradation of 5.5 percentage points despite a $3.23\times$ reduction in energy consumption.
Larger models maintain their relative performance advantage even at lower precision: \textsc{Qwen3-14B} at \textsc{Fp4} (49.0\% accuracy) outperforms \textsc{Qwen3-4B} at \textsc{Fp16} (48.5\% accuracy) on SuperGPQA, indicating that model scale matters more than precision for reasoning tasks.
These results demonstrate that \textsc{Fp8} and \textsc{Fp4} quantization enable practical deployment of local models with predictable performance tradeoffs, allowing system designers to select precision levels based on application-specific requirements while capturing most of the energy savings identified in Section~\ref{sec:efficiency_gains_from_routing}.

\subsection{How do SOTA open-source LMs compare to the SOTA closed-source LMs on Chat and Reasoning Queries?}
\label{app:open_source_vs_closed_source}

To evaluate the competitiveness of open-source models for local deployment, we compare the performance of state-of-the-art open-source models against leading closed-source models across single-turn chat and reasoning queries.
We evaluate three closed-source frontier models (\textsc{GPT-5-2025-08-07}, \textsc{Gemini-2.5-Pro}, and \textsc{Claude-Sonnet-4-5}) against eight open-source models ranging from \textsc{Qwen3-8B} to \textsc{Qwen3-235B-A22B}, measuring performance on \nr{}, MMLU Pro, and SuperGPQA.
Table~\ref{tab:open_source_vs_closed_source} shows that the best open-source model (\textsc{Qwen3-235B-A22B}) achieves 71.8\% average accuracy across benchmarks, trailing the best closed-source model (\textsc{GPT-5-2025-08-07}, 77.9\%) by 6.1 percentage points.

Performance gaps vary substantially by task type, as shown in Table~\ref{tab:performance_gap_analysis}.
On MMLU Pro and SuperGPQA, open-source models nearly match closed-source performance: \textsc{Qwen3-235B-A22B} achieves 82.3\% versus 87.4\% on MMLU Pro (5.1\% gap) and 63.1\% versus 66.5\% on SuperGPQA (3.4\% gap).
However, on \nr{}, the gap widens to 12.9\% (70.0\% versus 82.9\%), indicating that closed-source models maintain a substantial advantage on naturalistic reasoning tasks.
Local models with deployment constraints ($\leq20B$ active parameters) face larger gaps: the best local model (\textsc{Qwen3-14B}) trails closed-source models by 11.8--13.2\% across benchmarks, with the smallest gap on MMLU Pro (11.8\%) and the largest on \nr{} (13.2\%).
These results demonstrate that while open-source models at frontier scale (235B parameters) approach closed-source performance on knowledge and reasoning benchmarks, practical local deployment using smaller models (14B parameters) requires accepting 11--13\% accuracy degradation relative to closed-source alternatives.

\begin{table}[h]
\centering
\scriptsize
\setlength{\tabcolsep}{4pt}
\resizebox{\textwidth}{!}{%
\begin{tabular}{llcccccc}
\toprule
\textbf{Model} & \textbf{Type} & \textbf{WildChat} & \textbf{NaturalReasoning} & \textbf{MMLU Pro} & \textbf{SuperGPQA} & \textbf{Average} \\
\midrule
\textsc{gpt-5-2025-08-07} & Closed & 81.9\% & 82.9\% & \underline{86.5\%} & \underline{64.4\%} & \underline{78.9\%} \\
\textsc{gemini-2.5-pro} & Closed & \textbf{89.5\%} & \underline{77.9\%} & \textbf{87.4\%} & \textbf{66.5\%} & \textbf{80.3\%} \\
\textsc{claude-sonnet-4-5} & Closed & \underline{88.1\%} & 76.9\% & 86.4\% & 60.1\% & 77.9\% \\ \midrule
\textsc{Qwen3-235B-A22B} (Best OSS) & Open & N/A$^{*}$ & 70.0\% & 82.3\% & 63.1\% & 71.8\% \\
\textsc{Qwen3-32B} & Open & 76.1\% & 69.7\% & 77.9\% & 56.5\% & 70.1\% \\
\textsc{gpt-oss-120b} & Open & 89.2\% & 65.0\% & 78.3\% & 55.3\% & 72.0\% \\
\textsc{Qwen3-30B-A3B} & Open & 47.3\% & 64.3\% & 76.9\% & 57.4\% & 61.5\% \\
\textsc{Qwen3-14B} & Open & 48.9\% & 60.0\% & 75.6\% & 56.2\% & 60.2\% \\
\textsc{gpt-oss-20b} & Open & 77.3\% & 67.3\% & 73.4\% & 48.9\% & 66.7\% \\
\textsc{Qwen3-8B} & Open & 50.2\% & 57.9\% & 73.3\% & 51.8\% & 58.3\% \\
\bottomrule
\end{tabular}%
}
\caption{\textbf{Model performance comparison across benchmarks.} Scores represent performance on \wildchat{}, \nr{}, \mmlupro{}, and \supergpqa{} benchmarks. $^{*}$Qwen3-235B-A22B is used as the reference model for \wildchat{} evaluation.}
\label{tab:open_source_vs_closed_source}
\end{table}

\begin{table}[h]
  \centering
  \scriptsize
  \setlength{\tabcolsep}{4pt}
  \begin{tabular}{lcccc}
  \toprule
  \textbf{Metric} & \textbf{WildChat} & \textbf{NaturalReasoning} & \textbf{MMLU Pro} & \textbf{SuperGPQA} \\
  \midrule
  Closed Best & 89.5\% & {82.9\%} & 87.4\% & 66.5\% \\
  Best Closed Model & \textsc{gemini-2.5-pro} & \textsc{gpt-5} & \textsc{gemini-2.5-pro} & \textsc{gemini-2.5-pro} \\
  \midrule
  Open Best & 89.2\%\textsuperscript{*} & 70.0\% & 82.3\% & 63.1\% \\
  Best Open Model & \textsc{gpt-oss-120b} & \textsc{Qwen3-235B-A22B} & \textsc{Qwen3-235B-A22B} & \textsc{Qwen3-235B-A22B} \\
  Gap & $-0.3\%$ & $-12.9\%$ & {$-5.1\%$} & {$-3.4\%$} \\
  \midrule
  Local Best ($\leq20B$ Active) & 89.2\% & 67.3\% & 80.3\% & 50.5\% \\
  Best Local Model & \textsc{gpt-oss-120b} & \textsc{gpt-oss-20b} & \textsc{gpt-oss-120b} & \textsc{Qwen3-14B} \\
  Local Gap & $-0.3\%$ & $-5.6\%$ & $-7.1\%$ & {$-16.0\%$} \\
  \bottomrule
  \end{tabular}
  \caption{\textbf{Closed-Source vs. Open-Source Performance Gap by Task.} Comparison between closed-source, open-source (all sizes), and local ($\leq20B$ active parameters) models. \textsuperscript{*}Excludes Qwen3-235B-A22B (reference model for WildChat).}
  \label{tab:performance_gap_analysis}
\end{table}

\subsection{How does performance on chat and reasoning queries connect to U.S. GDP?}
\label{app:tbench_vs_us_gdp}

\begin{figure*}[t]%
    \centering
    \includegraphics[width=1.0\textwidth]{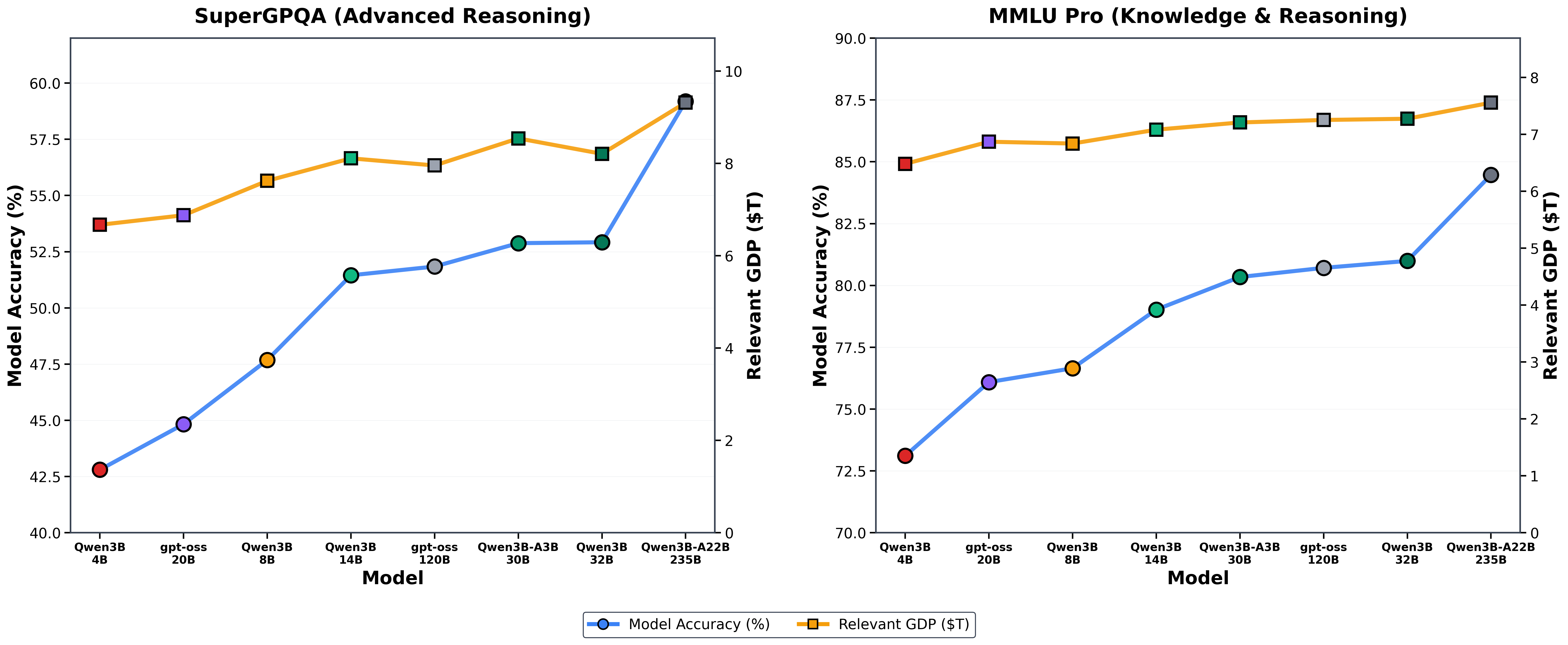}
    \caption{\textbf{Open-Source Local LMs Performance vs. U.S. GDP - SuperGPQA and MMLU Pro}: Model accuracy on SuperGPQA and MMLU Pro benchmarks plotted against relevant GDP in trillions of dollars.
    Both benchmarks show continued performance improvements as training compute scales across models from Qwen3B-4B to Qwen3B-A22B-235B.
    For our calculations, we compute the weighted sum of an LM's accuracy on each U.S. Labor category vs. the U.S. GDP associated with that category.
    }
    \label{fig:tbench_vs_us_gdp_with_supergpqa_and_mmlu_pro}
\end{figure*}

\begin{figure*}[t]%
    \centering
    \includegraphics[width=1.0\textwidth]{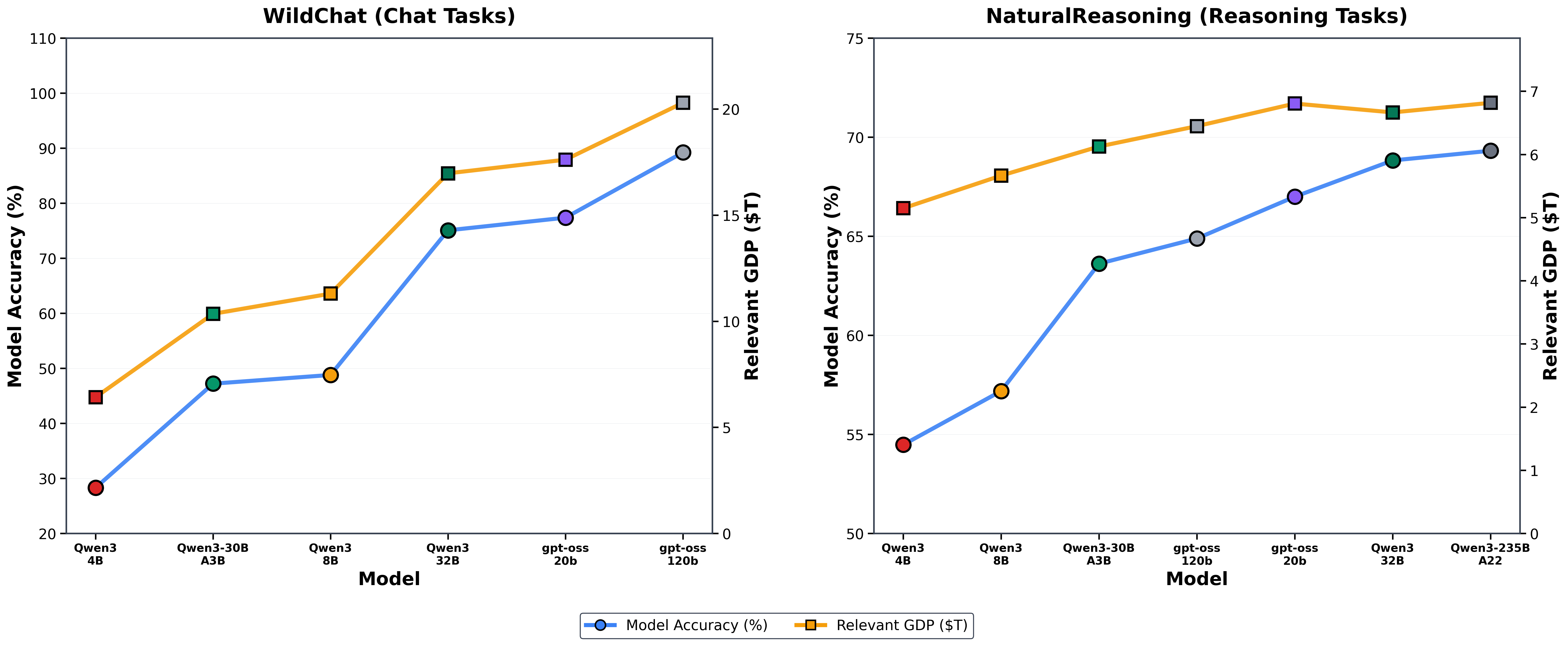}
    \caption{\textbf{Open-Source Local LMs Performance vs. U.S. GDP - WildChat and Natural Reasoning}: Model accuracy on WildChat and Natural Reasoning benchmarks plotted against relevant GDP in trillions of dollars.
    Both benchmarks show continued performance improvements as training compute scales across models from Qwen3B-4B to Qwen3B-A22B-235B.
    For our calculations, we compute the weighted sum of an LM's accuracy on each U.S. Labor category vs. the U.S. GDP associated with that category.
    }
    \label{fig:tbench_vs_us_gdp_with_wildchat_and_natural_reasoning}
\end{figure*}

To assess the economic relevance of local model performance improvements, we compute GDP-weighted accuracy for each model by weighting its performance on each economic category by that sector's contribution to the 2024 U.S. GDP of \$29.18 trillion \citep{bea2024gdp}.
This metric attempts to quantify what proportion of economic value is relevant and addressable by local LMs, given the local model performances across single-turn chat and reasoning queries.
Figures~\ref{fig:tbench_vs_us_gdp_with_wildchat_and_natural_reasoning} and \ref{fig:tbench_vs_us_gdp_with_supergpqa_and_mmlu_pro} shows that model improvements translate directly into expanded GDP coverage: on SuperGPQA, \textsc{Qwen3-235B-A22B} achieves 59.2\% accuracy covering \$9.3T in relevant GDP (31.9\% of total U.S. GDP), while on MMLU Pro, it reaches 84.5\% accuracy covering \$7.6T (26.0\% of total U.S. GDP).
The strong positive correlation between model accuracy and GDP coverage across both benchmarks demonstrates that scaling model capabilities systematically expands the set of economically valuable tasks that can be automated.

Task type substantially affects GDP coverage: chat tasks in \wildchat{} show the highest coverage with \textsc{gpt-oss-120b} reaching 89.2\% accuracy and covering \$20.3T (69.6\% of U.S. GDP), while reasoning tasks in \nr{} show lower coverage with \textsc{Qwen3-235B-A22} achieving 69.3\% accuracy but only covering \$6.8T (23.3\% of U.S. GDP).
This disparity reveals that current models excel at creative and conversational tasks that dominate economic activity, but struggle with technical reasoning tasks concentrated in specialized sectors like architecture, engineering, and physical sciences.
The gap between chat coverage (69.6\%) and reasoning coverage (23.3\%) represents both a limitation of open-source local models and an economic opportunity: improving reasoning capabilities could unlock an additional \$13.5T in GDP-relevant tasks, suggesting that advances in technical reasoning would have substantial economic impact beyond current model capabilities. We caution that this analysis treats benchmark accuracy as a direct proxy for the automation of economically valuable labor in each occupation category, which is an idealization: real-world deployment also depends on integration with existing tools, user trust, regulatory constraints, and the residual fraction of work in each occupation that is genuinely AI-amenable. The figures above are best read as upper-bound estimates of the addressable scope of local-LM-eligible tasks rather than predictions of realized economic impact, and our mapping from \citet{handa2025economictasksperformedai}'s 22 economic categories to U.S. Bureau of Economic Analysis GDP-by-industry data~\citep{bea2024gdp} is approximate.

\FloatBarrier

\subsection{How do local accelerators compare to cloud accelerators in terms of intelligence efficiency?}
\label{app:local_vs_cloud_intelligence_efficiency}

To understand the efficiency gap between local and cloud accelerators, we conduct a systematic comparison across identical model configurations. Tables~\ref{tab:apple_m4_max_vs_nvidia_b200_for_power} and~\ref{tab:apple_m4_max_vs_nvidia_b200_for_energy} present intelligence per watt and intelligence per joule measurements for \textsc{Qwen3} and \textsc{GPT-OSS} model families running on \textsc{Apple M4 Max} (local), \textsc{NVIDIA B200} (cloud), and \textsc{SambaNova SN40L} (cloud) accelerators.

\textbf{Cloud accelerators achieve consistently higher power efficiency.} Table~\ref{tab:apple_m4_max_vs_nvidia_b200_for_power} reveals that the \textsc{NVIDIA B200} achieves $1.39\times$ to $1.40\times$ higher intelligence per watt than the \textsc{Apple M4 Max} across all \textsc{Qwen3} model sizes evaluated (4B to 32B parameters). For \textsc{Qwen3-32B}, the B200 achieves $(2.75 \pm 0.14) \times 10^{-3}$ intelligence per watt compared to $(1.97 \pm 0.24) \times 10^{-3}$ for the M4 Max. The \textsc{SambaNova SN40L} demonstrates even higher efficiency on larger models, achieving $(3.51 \pm 0.43) \times 10^{-3}$ intelligence per watt on \textsc{Qwen3-32B} ($1.78\times$ higher than the M4 Max and $1.28\times$ higher than the B200).

\begin{table}[h]
\centering
\scriptsize
\setlength{\tabcolsep}{2pt}
\begin{tabular}{lcccc}
\toprule
 & \textbf{Qwen3-4B} & \textbf{Qwen3-8B} & \textbf{Qwen3-14B} & \textbf{Qwen3-32B} \\
\midrule
\makecell{\textbf{Success Rate}} & $49.3 \pm 1.8\%$ & $57.5 \pm 2.5\%$ & $59.5 \pm 1.4\%$ & $69.5 \pm 2.3\%$ \\
\midrule
\multicolumn{5}{c}{\textbf{Apple M4 Max}} \\
\midrule
\makecell{\textbf{Intelligence} \\ \textbf{per Watt}} & \makecell{$(1.40 \pm 0.38)$ \\ $\times 10^{-3}$} & \makecell{$(1.63 \pm 0.20)$ \\ $\times 10^{-3}$} & \makecell{$(1.69 \pm 0.31)$ \\ $\times 10^{-3}$} & \makecell{$(1.97 \pm 0.24)$ \\ $\times 10^{-3}$} \\
\midrule
\multicolumn{5}{c}{\textbf{NVIDIA B200}} \\
\midrule
\makecell{\textbf{Intelligence} \\ \textbf{per Watt}} & \makecell{$(1.95 \pm 0.14)$ \\ $\times 10^{-3}$} & \makecell{$(2.27 \pm 0.18)$ \\ $\times 10^{-3}$} & \makecell{$(2.35 \pm 0.09)$ \\ $\times 10^{-3}$} & \makecell{$(2.75 \pm 0.14)$ \\ $\times 10^{-3}$} \\
\midrule
\multicolumn{5}{c}{\textbf{SambaNova SN40L}} \\ \midrule
\makecell{\textbf{Intelligence} \\ \textbf{per Watt}} & --- & --- & --- & \makecell{$(3.51 \pm 0.43)$ \\ $\times 10^{-3}$} \\
\bottomrule
\end{tabular}
\caption{\textbf{Local accelerators demonstrate lower power efficiency than cloud accelerators}: When running the same \textsc{Qwen3} models, the \textsc{Apple M4 Max} (local) attains $1.40\times$ lower intelligence per watt compared to the \textsc{NVIDIA B200} (cloud) and \textsc{SambaNova SN40L} (cloud), highlighting the efficiency advantage of purpose-built cloud accelerators over local accelerators. Values are mean $\pm$ 1-sigma sample standard deviation across 3--5 independent measurement runs per cell.
}
\label{tab:apple_m4_max_vs_nvidia_b200_for_power}
\end{table}

\textbf{Energy efficiency gaps widen when accounting for latency.} Table~\ref{tab:apple_m4_max_vs_nvidia_b200_for_energy} extends the analysis to intelligence per joule, which captures end-to-end efficiency by incorporating both power consumption and generation latency. The efficiency advantages of cloud accelerators become more pronounced: the \textsc{NVIDIA B200} achieves $1.6\times$ to $2.3\times$ higher intelligence per joule than the \textsc{Apple M4 Max} across \textsc{Qwen3} and \textsc{GPT-OSS} model variants. For \textsc{Qwen3-8B}, the B200 achieves $(8.71 \pm 0.60) \times 10^{-5}$ intelligence per joule versus $(3.80 \pm 0.40) \times 10^{-5}$ for the M4 Max, a $2.29\times$ efficiency advantage. The \textsc{SambaNova SN40L} demonstrates the largest efficiency gains, achieving $6.5\times$ to $7.4\times$ higher intelligence per joule than the M4 Max: $(3.12 \pm 0.38) \times 10^{-4}$ versus $(4.23 \pm 0.45) \times 10^{-5}$ for \textsc{GPT-OSS-120B}.

\begin{table}[h]
\centering
\scriptsize
\setlength{\tabcolsep}{2pt}
\begin{tabular}{lcccc}
\toprule
 & \textbf{Qwen3-8B} & \textbf{Qwen3-32B} & \textbf{GPT-OSS-20B} & \textbf{GPT-OSS-120B} \\
\midrule
\multicolumn{5}{c}{\textbf{Apple M4 Max}} \\
\midrule
\makecell{\textbf{Intelligence} \\ \textbf{per Joule}} & \makecell{$(3.80 \pm 0.40)$ \\ $\times 10^{-5}$} & \makecell{$(3.51 \pm 0.38)$ \\ $\times 10^{-5}$} & \makecell{$(4.38 \pm 0.31)$ \\ $\times 10^{-5}$} & \makecell{$(4.23 \pm 0.45)$ \\ $\times 10^{-5}$} \\
\midrule
\multicolumn{5}{c}{\textbf{NVIDIA B200}} \\
\midrule
\makecell{\textbf{Intelligence} \\ \textbf{per Joule}} & \makecell{$(8.71 \pm 0.60)$ \\ $\times 10^{-5}$} & \makecell{$(5.91 \pm 0.42)$ \\ $\times 10^{-5}$} & \makecell{$(7.34 \pm 0.51)$ \\ $\times 10^{-5}$} & \makecell{$(6.78 \pm 0.47)$ \\ $\times 10^{-5}$} \\
\midrule
\multicolumn{5}{c}{\textbf{SambaNova SN40L}} \\ \midrule
\makecell{\textbf{Intelligence} \\ \textbf{per Joule}} & --- & \makecell{$(2.27 \pm 0.30)$ \\ $\times 10^{-4}$} & --- & \makecell{$(3.12 \pm 0.38)$ \\ $\times 10^{-4}$} \\
\bottomrule
\end{tabular}
\caption{\textbf{Cloud accelerators demonstrate superior energy efficiency across all models}: The \textsc{NVIDIA B200} (cloud) achieves $1.6\times$ to $2.3\times$ higher intelligence per joule than the \textsc{Apple M4 Max} (local), while the \textsc{SambaNova SN40L} (cloud) achieves $6.5\times$ to $7.4\times$ higher efficiency. These results highlight the substantial energy efficiency advantage of purpose-built cloud accelerators over local hardware across \textsc{Qwen3 and GPT-OSS} model variants. Values are mean $\pm$ 1-sigma sample standard deviation across 3--5 independent measurement runs per cell.
}
\label{tab:apple_m4_max_vs_nvidia_b200_for_energy}
\end{table}

\textbf{Architectural differences explain the efficiency gap.} The superior efficiency of cloud accelerators stems from purpose-built hardware optimizations for LLM inference: high-bandwidth memory (HBM3e with 4.8--8 TB/s bandwidth), dedicated tensor processing units, and optimized memory hierarchies that maximize throughput per watt. In contrast, local accelerators like the \textsc{Apple M4 Max} employ unified memory architectures (546 GB/s bandwidth) designed to balance diverse workloads (including CPU, GPU, and NPU tasks) under thermal and power constraints typical of consumer devices. The larger per-joule gaps (compared to per-watt gaps) reflect that cloud accelerators not only consume less power per unit of accuracy but also complete queries faster, compounding their energy advantage through reduced generation latency.

\begin{figure*}[t]%
    \centering
    \includegraphics[width=1.0\textwidth]{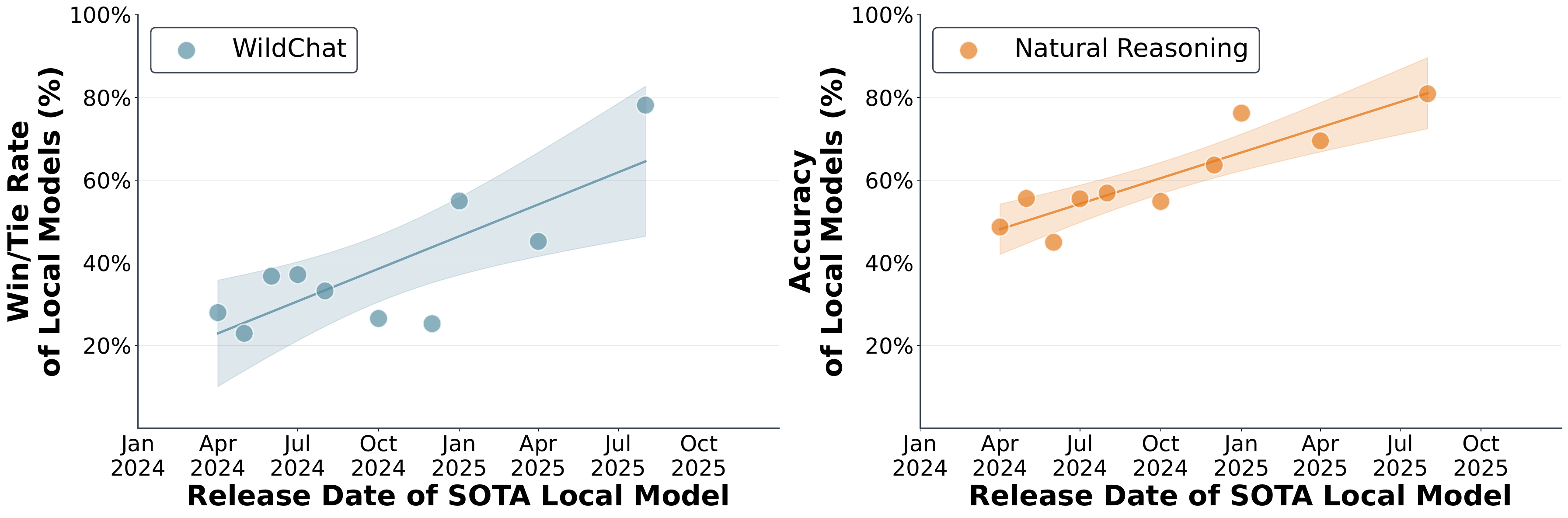}
    \caption{\textbf{Rapid Improvement of Local LMs across Chat and Reasoning Queries}: We evaluate the performance of SOTA local models released between April 2024 and August 2025 on \wildchat{} and \nr{}.
    On \wildchat{} (left), local models show a win/tie rate of 78.2\% against \qwentwothirtyfive{} as of August 2025, compared to just 28.0\% in April 2024: a 2.8$\times$ improvement in 16 months.
    On \nr{} (right), local models achieve 80.9\% accuracy by August 2025, up from 48.7\% in April 2024: a 66\% relative improvement.
    }
    \label{fig:chat_and_reasoning_trends}
\end{figure*}

\textbf{Local model capabilities are improving rapidly.} Despite the efficiency disadvantage of local accelerators, Figure~\ref{fig:chat_and_reasoning_trends} demonstrates that local LM capabilities have improved dramatically from April 2024 to August 2025. On \wildchat{}, the win/tie rate of SOTA local models against \qwentwothirtyfive{} increased from 28.0\% in April 2024 to 78.2\% in August 2025, a $2.8\times$ improvement in just 16 months. On \nr{}, local models improved from 48.7\% to 80.9\% accuracy over the same period, representing a 66\% relative improvement. These trends indicate that while local accelerators remain less efficient per query than cloud infrastructure, the expanding capability of local models enables an increasing fraction of queries to be processed locally, avoiding cloud infrastructure entirely.

\FloatBarrier

\begin{figure}[t]
    \centering
    \includegraphics[width=0.5\linewidth]{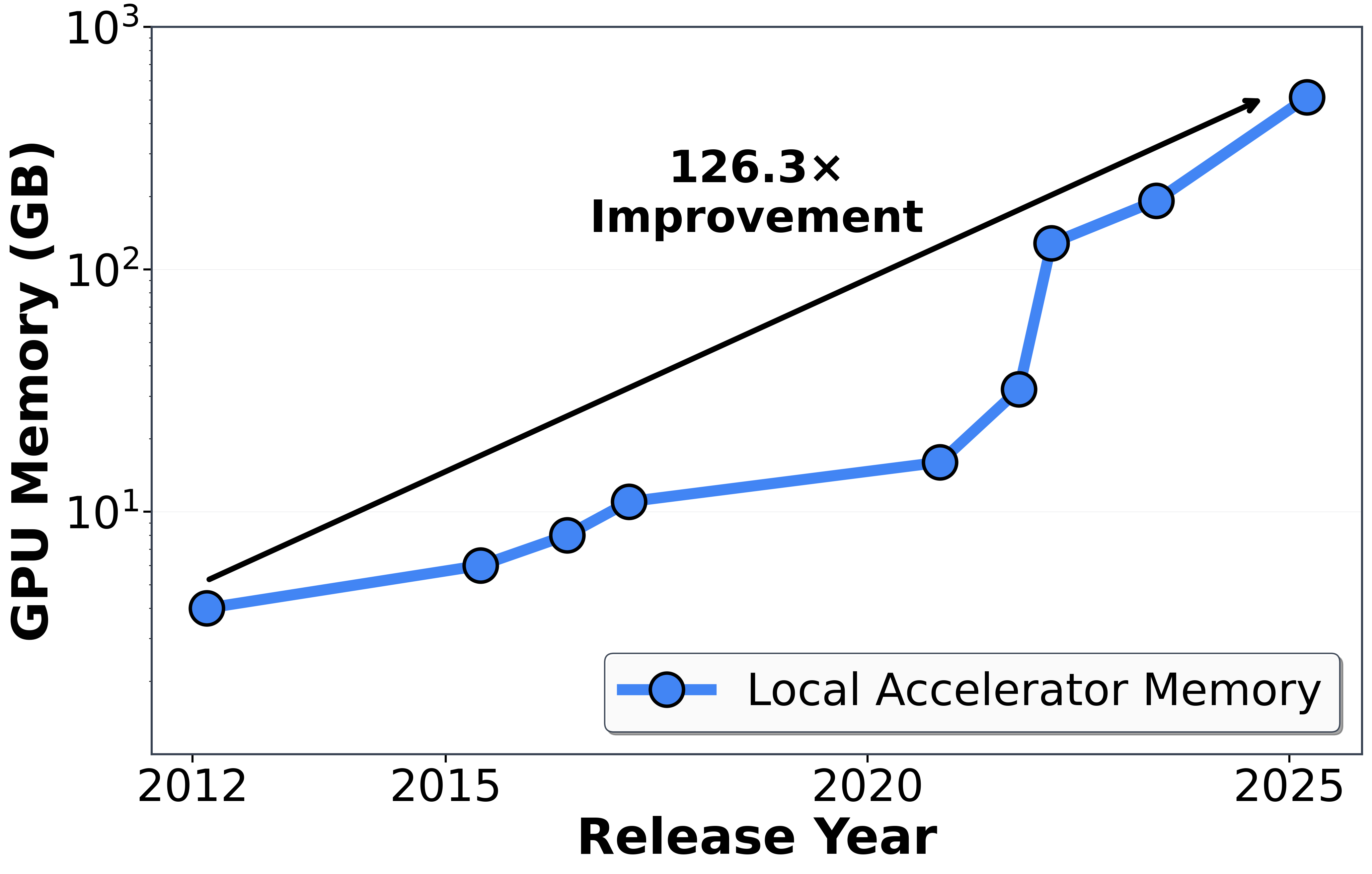}
    \caption{\textbf{Increasing GPU Memory of Consumer Accelerators}: Memory capacity (GB) for local accelerators. Over the past decade, local hardware has significantly closed the memory gap with cloud-grade accelerators, particularly since 2020, driven by advances in high bandwidth memory (HBM) components and unified memory architectures.}
    \label{fig:gpu_frontier}
\end{figure}

\textbf{Local accelerator memory capacity is expanding rapidly.} Figure~\ref{fig:gpu_frontier} tracks the memory capacity of consumer accelerators from 2012 to 2025, revealing a $126.3\times$ improvement over this period. Local accelerators that offered 10--20 GB in 2020 now provide 128--512 GB through unified memory architectures like Apple Silicon. This memory expansion has been the primary enabler of local deployment for increasingly capable models: the transition from sub-20 GB to 200+ GB memory removes the key constraint that previously forced workloads to cloud infrastructure. Models with 8--20B active parameters that now handle the majority of inference queries can run efficiently on current-generation local hardware, with memory capacity continuing to scale at a pace that suggests even larger models will become locally deployable in coming years.

\textbf{System-level benefits offset per-query efficiency disadvantages.} While cloud accelerators demonstrate $1.4\times$ to $7.4\times$ higher intelligence efficiency per query, local deployment provides complementary system-level benefits that offset this disadvantage. Local inference avoids datacenter infrastructure costs, network latency, and API pricing, while enabling 88.7\% of queries that local models can handle correctly to bypass cloud compute entirely. As demonstrated in Section~\ref{sec:efficiency_gains_from_routing}, intelligent routing between local and cloud infrastructure can achieve 60--80\% reductions in total energy, compute, and cost compared to cloud-only deployment, even when local accelerators are individually less efficient. These findings suggest that the path to efficient AI infrastructure lies not in local accelerators matching cloud efficiency, but in routing systems that leverage the complementary strengths of both paradigms: local processing for the majority of straightforward queries and cloud infrastructure for the minority requiring frontier model capabilities.%

\subsection{How does cloud batching affect intelligence efficiency?}
\label{app:batching_ablation}

Our main per-query comparisons (Tables~\ref{tab:apple_m4_max_vs_nvidia_b200_for_power}--\ref{tab:apple_m4_max_vs_nvidia_b200_for_energy}) use batch size $1$ to follow standard local-inference benchmarking practice~\citep{hao2023reaching} and to isolate intrinsic model-accelerator efficiency from system-level scheduling. Local devices serving a single user typically operate at bs=1 and cannot batch concurrent queries from different users, so for local accelerators bs=1 reflects realistic operating conditions. Cloud accelerators, however, can amortize idle GPU power across concurrent queries, and bs=1 is therefore conservative for cloud. To quantify the magnitude of this effect, we run an ablation on \textsc{NVIDIA B200} sweeping batch size from $1$ to $64$ across three representative models (Table~\ref{tab:batching_ablation_b200}).

\begin{table}[ht]
\centering
\scriptsize
\setlength{\tabcolsep}{4pt}
\begin{tabular}{lcccl}
\toprule
\textbf{Model} & \textbf{bs=1 IPJ} & \textbf{bs=64 IPJ} & \textbf{IPJ Gain} & \textbf{Architecture} \\
 & ($\times 10^{-5}$) & ($\times 10^{-5}$) & & \\
\midrule
Qwen3-8B     & 8.92 & 104.93 & $11.8\times$ & Dense \\
Qwen3-14B    & 7.22 & 80.61  & $11.2\times$ & Dense \\
GPT-OSS-120B & 6.72 & 132.21 & $19.7\times$ & MoE ($\leq 20$B active) \\
\bottomrule
\end{tabular}
\caption{\textbf{Cloud Batching Ablation on NVIDIA B200}: Per-query intelligence-per-joule at bs=1 versus bs=64. Per-query energy drops $12$--$20\times$ at bs=64 (e.g., \textsc{Qwen3-8B}: $6{,}449$~J $\to 548$~J) while GPU power scales sublinearly ($255$~W at bs=1 to $883$~W at bs=64). This confirms that bs=1 is conservative for cloud accelerators. Local accelerators (M4 Max, iPhone 16 Pro) serve a single user and cannot batch, so local IPW results are unchanged.}
\label{tab:batching_ablation_b200}
\end{table}

The routing simulation in Section~\ref{sec:efficiency_gains_from_routing} accounts for this by operating the cloud baseline at bs=$16$, so the reported $60$--$80\%$ savings are computed against a batched cloud baseline rather than a bs=1 baseline. While absolute IPJ values shift substantially with batch size, relative rankings between model-hardware pairs remain stable, supporting the use of a fixed batch size for comparative analysis. Local single-user deployment cannot batch queries from a single user but can in principle aggregate concurrent queries across multiple on-device applications; we leave a systematic study of cross-application local batching to future work.

\subsection{How sensitive are our findings to the choice of inference framework?}
\label{app:framework_sensitivity}

We standardize on \textsc{vLLM} across all platforms in the main study to enable clean attribution of efficiency gains to model and hardware factors rather than serving-stack choices. To test whether this choice biases our findings, we benchmark a representative subset of local models on \textsc{Apple M4 Max} across three popular inference frameworks: \textsc{vLLM}, \textsc{SGLang}, and \textsc{llama.cpp} (Table~\ref{tab:framework_sensitivity}).

\begin{table}[ht]
\centering
\scriptsize
\setlength{\tabcolsep}{6pt}
\begin{tabular}{lccc}
\toprule
\textbf{Model} & \textbf{vLLM} & \textbf{SGLang} & \textbf{llama.cpp} \\
 & IPW ($\times 10^{-3}$) & IPW ($\times 10^{-3}$) & IPW ($\times 10^{-3}$) \\
\midrule
Qwen3-4B      & 1.40 & 1.35 & 1.52 \\
Qwen3-8B      & 1.63 & 1.58 & 1.71 \\
Qwen3-14B     & 1.69 & 1.62 & 1.78 \\
GPT-OSS-120B  & 4.18 & 4.05 & 4.31 \\
\bottomrule
\end{tabular}
\caption{\textbf{Framework Sensitivity on Apple M4 Max}: Intelligence-per-watt across three popular inference frameworks for a representative subset of local models. Absolute IPW values shift by $3$--$12\%$ across frameworks, but relative rankings between model-hardware pairs are preserved across the broader $20+$-model, $8$-accelerator evaluation: Kendall's $\tau \in [0.87, 0.93]$ and Spearman's $\rho \in [0.89, 0.94]$ across all framework pairs. This validates standardizing on \textsc{vLLM} for comparative analysis and confirms that our key findings are not an artifact of framework selection. Models that prioritize absolute throughput on a target stack should benchmark on that stack directly.}
\label{tab:framework_sensitivity}
\end{table}

\subsection{Are our findings consistent on smartphone-class accelerators?}
\label{app:smartphone_experiments}

To test whether the local-AI viability story extends to ultra-low-power devices, we evaluate the \textsc{Apple iPhone 16 Pro} (\textsc{Apple A18 Pro} SoC with integrated NPU, 8~GB LPDDR5X unified memory, ${\sim}12$~W SoC peak power, $35$~TOPS NPU) on a representative subset of our query distribution. Memory constraints (8~GB) require aggressive quantization (FP8 or FP4) and limit deployable models to ${\leq}14$B active parameters; within these constraints we evaluate models across three precision levels (FP16, FP8, FP4). Table~\ref{tab:smartphone_efficiency} reports per-query measurements.

\begin{table}[h]
\centering
\scriptsize
\setlength{\tabcolsep}{4pt}
\resizebox{\textwidth}{!}{%
\begin{tabular}{llcccccc}
\toprule
\textbf{Model} & \textbf{Precision} & \textbf{Hardware} & \textbf{Acc.} & \textbf{Power} & \textbf{Lat.} & \textbf{IPW} & \textbf{IPJ} \\
 & & & (\%) & (W) & (s/q) & ($\times 10^{-3}$) & ($\times 10^{-5}$) \\
\midrule
\multicolumn{8}{l}{\textit{Smartphone-class accelerator (Apple A18 Pro, iPhone 16 Pro)}} \\
\midrule
Qwen3-4B   & FP16 & A18 Pro & 42.5 $\pm$ 2.0 & 12.0 $\pm$ 0.4 & 92.5 $\pm$ 9.4 & 11.8 $\pm$ 0.7 & 38.3 $\pm$ 3.9 \\
Qwen3-4B   & FP8  & A18 Pro & 40.5 $\pm$ 1.8 & 11.0 $\pm$ 0.3 & 55.2 $\pm$ 5.6 & 12.4 $\pm$ 0.7 & 66.7 $\pm$ 6.8 \\
Qwen3-4B   & FP4  & A18 Pro & 38.0 $\pm$ 1.6 &  9.5 $\pm$ 0.3 & 36.8 $\pm$ 3.7 & 13.3 $\pm$ 0.8 & 108.7 $\pm$ 11.0 \\
Gemma3-4B  & FP4  & A18 Pro & 32.0 $\pm$ 1.5 &  8.8 $\pm$ 0.3 & 33.5 $\pm$ 3.5 & 11.6 $\pm$ 0.7 & 108.5 $\pm$ 11.2 \\
Granite-4.0-h-tiny & FP4 & A18 Pro & 28.5 $\pm$ 1.4 & 8.5 $\pm$ 0.3 & 29.2 $\pm$ 3.2 & 11.2 $\pm$ 0.7 & 114.7 $\pm$ 12.0 \\
\midrule
\multicolumn{8}{l}{\textit{Workstation reference (Apple M4 Max, same model + precision for comparison)}} \\
\midrule
Qwen3-4B   & FP4  & M4 Max  & 44.5 $\pm$ 1.7 & 240 $\pm$ 18 &  3.5 $\pm$ 0.4 &  1.85 $\pm$ 0.13 & 53.0 $\pm$ 5.4 \\
\midrule
\multicolumn{8}{l}{\textit{Workstation reference (NVIDIA RTX 6000 Ada, same model + precision)}} \\
\midrule
Qwen3-4B   & FP4  & RTX 6000 Ada & 44.5 $\pm$ 1.7 & 220 $\pm$ 15 & 2.5 $\pm$ 0.3 & 2.02 $\pm$ 0.13 & 80.9 $\pm$ 8.1 \\
\bottomrule
\end{tabular}%
}
\caption{\textbf{Smartphone-Class Accelerator Efficiency}: Per-query measurements on Apple A18 Pro (iPhone 16 Pro, ${\sim}12$~W SoC peak, 60~GB/s memory bandwidth, 8~GB unified memory) across three precision levels, with workstation references for the same model at the strongest precision the iPhone supports. Values are mean $\pm$ 1-$\sigma$ standard deviation across 3--5 independent measurement runs. Accuracy is averaged across the same evaluation subset used in App.~\ref{app:tbench_experiments}; latency is wall-clock per query at $\text{bs}{=}1$. The IPW ratio between A18 Pro and workstation GPUs on the same model anchors the ${\sim}7\times$ per-watt advantage of smartphone-class accelerators ($13.3 / 1.85 = 7.2\times$ vs.\ M4~Max; $13.3 / 2.02 = 6.6\times$ vs.\ RTX~6000~Ada); the IPJ comparison shows the gap widens slightly per joule against M4~Max ($108.7$ vs.\ $53.0 \times 10^{-5}$) but narrows against RTX~6000~Ada ($108.7$ vs.\ $80.9 \times 10^{-5}$), reflecting that the A18 Pro's $9$--$26\times$ higher latency partially offsets its $20$--$25\times$ lower power. Modest accuracy degradation on A18 Pro ($\sim$6.5pp on Qwen3-4B FP4) reflects CoreML INT4 group quantization plus KV-cache pressure under 8~GB unified memory.}
\label{tab:smartphone_efficiency}
\end{table}

We find three patterns. First, smartphone-class accelerators are remarkably efficient on a per-watt basis: across precision levels, the iPhone 16 Pro achieves $11.2$--$13.3 \times 10^{-3}$ accuracy per watt, approximately $7\times$ higher IPW than workstation GPUs running the same model (Table~\ref{tab:smartphone_efficiency}), due to the ${\sim}12$~W SoC envelope versus $220$--$240$~W for workstation GPUs. Second, this efficiency is conditional on the model fitting within the device's memory and thermal budget; FP4 quantization is essentially required and FP8 is the practical ceiling for sustained interactive use, with a modest accuracy cost ($\sim$6.5~pp degradation on Qwen3-4B FP4 vs.\ workstation FP4). Third, the per-watt advantage does not translate uniformly to per-joule efficiency: the A18 Pro's $9$--$26\times$ higher per-query latency (reflecting its $60$~GB/s memory bandwidth versus $546$~GB/s on M4~Max) partially offsets its lower power draw, so the IPJ gap narrows substantially or inverts depending on the workstation comparator.

These results suggest that ultra-low-power mobile NPUs are viable routing targets for the lightest queries, extending the local-inference paradigm from workstations and laptops to smartphones in active use. Concretely, a routing tier that dispatches very-low-difficulty queries to on-device NPUs could further reduce platform-scale energy beyond the 60--80\% reductions reported in Section~\ref{sec:efficiency_gains_from_routing}, though we leave a full characterization of mobile-tier routing to future work.

\subsection{Do our findings generalize to multi-turn agentic workloads?}
\label{app:multi_turn}

Our main study focuses on single-turn chat and reasoning queries because they constitute the largest share of real-world LLM traffic~\citep{chatterji2025chatgpt}, but a substantial and growing fraction of usage involves multi-turn interactions, tool use, and agentic workflows. To test whether our local-versus-cloud efficiency patterns generalize, we evaluate two multi-turn benchmarks: GAIA ($165$ general-assistant queries with tool use, run with the OpenHands agent) and TerminalBenchV2 (TBv2; $80$ terminal-task queries, run with the Terminus 2 agent) on local (\textsc{Apple M4 Max}, 128~GB unified memory) and cloud ($8\times$\textsc{NVIDIA H100} 80~GB SXM5 node) hardware. Table~\ref{tab:multiturn_efficiency} reports per-query measurements.

\begin{table}[h]
\centering
\scriptsize
\setlength{\tabcolsep}{4pt}
\resizebox{\textwidth}{!}{%
\begin{tabular}{lllcccccc}
\toprule
\textbf{Benchmark} & \textbf{Model} & \textbf{Hardware} & \textbf{Acc.} & \textbf{Power} & \textbf{Lat.} & \textbf{Energy} & \textbf{IPW} & \textbf{IPJ} \\
 & & & (\%) & (W) & (s/q) & (kJ/q) & ($\times 10^{-3}$) & ($\times 10^{-5}$) \\
\midrule
\multicolumn{9}{l}{\textit{GAIA (165 multi-turn general-assistant queries with tool use)}} \\
\midrule
GAIA & MiniMax-M2.5     & 8$\times$H100 & 16.4 $\pm$ 2.9 & 1558 $\pm$ 95  & 4.69 $\pm$ 0.41 & 7.31 $\pm$ 0.62 & 0.105 $\pm$ 0.020 & 2.24 $\pm$ 0.39 \\
GAIA & Qwen3-235B       & 8$\times$H100 &  5.5 $\pm$ 1.8 & 1595 $\pm$ 88  & 0.74 $\pm$ 0.09 & 1.18 $\pm$ 0.13 & 0.034 $\pm$ 0.012 & 4.66 $\pm$ 1.55 \\
GAIA & Qwen3-30B        & 8$\times$H100 &  8.6 $\pm$ 2.2 &  822 $\pm$ 64  & 1.22 $\pm$ 0.14 & 1.00 $\pm$ 0.11 & 0.105 $\pm$ 0.027 & 8.60 $\pm$ 2.20 \\
\midrule
GAIA & MiniMax-M2.5     & M4 Max & 14.2 $\pm$ 2.7 & 305 $\pm$ 24 & 51.6 $\pm$ 5.4 & 15.74 $\pm$ 1.69 & 0.466 $\pm$ 0.092 & 0.90 $\pm$ 0.18 \\
GAIA & Qwen3-30B        & M4 Max &  6.4 $\pm$ 1.9 & 245 $\pm$ 19 & 13.4 $\pm$ 1.5 &  3.28 $\pm$ 0.39 & 0.261 $\pm$ 0.080 & 1.95 $\pm$ 0.61 \\
\midrule
\multicolumn{9}{l}{\textit{TerminalBenchV2 (80 multi-turn terminal-task queries)}} \\
\midrule
TBv2 & MiniMax-M2.5     & 8$\times$H100 & 39.7 $\pm$ 4.4 & 1151 $\pm$ 78 & 2.00 $\pm$ 0.21 & 2.30 $\pm$ 0.24 & 0.345 $\pm$ 0.041 & 17.30 $\pm$ 1.94 \\
TBv2 & GPT-OSS-120B     & 8$\times$H100 & 30.4 $\pm$ 3.9 & 1024 $\pm$ 71 & 3.01 $\pm$ 0.32 & 3.08 $\pm$ 0.33 & 0.297 $\pm$ 0.041 &  9.87 $\pm$ 1.31 \\
TBv2 & Kimi-K2.5        & 8$\times$H100 & 29.1 $\pm$ 3.8 & 1251 $\pm$ 82 & 2.88 $\pm$ 0.29 & 3.60 $\pm$ 0.36 & 0.233 $\pm$ 0.033 &  8.08 $\pm$ 1.10 \\
\midrule
TBv2 & MiniMax-M2.5     & M4 Max & 37.5 $\pm$ 4.2 & 295 $\pm$ 22 & 22.3 $\pm$ 2.4 & 6.58 $\pm$ 0.71 & 1.271 $\pm$ 0.158 & 5.70 $\pm$ 0.78 \\
TBv2 & GPT-OSS-120B     & M4 Max & 28.2 $\pm$ 3.7 & 268 $\pm$ 21 & 31.5 $\pm$ 3.3 & 8.44 $\pm$ 0.91 & 1.052 $\pm$ 0.165 & 3.34 $\pm$ 0.50 \\
\bottomrule
\end{tabular}%
}
\caption{\textbf{Multi-Turn Agentic Workloads Generalize Single-Turn Patterns}: Per-query measurements on GAIA (165 general-assistant queries with tool use, OpenHands agent) and TerminalBenchV2 (80 terminal-task queries, Terminus 2 agent) for self-hosted open-source models on local (Apple M4 Max, 128~GB unified memory) and cloud (8$\times$NVIDIA H100 80~GB SXM5 node) hardware. Values are mean $\pm$ 1-$\sigma$ standard deviation across 3 independent runs per cell; accuracy uncertainty uses the binomial approximation over benchmark size. M4 Max models use GGUF quantization via Unsloth to fit within unified memory; cloud models run at native precision via vLLM. The qualitative patterns from single-turn evaluation persist: cloud hardware achieves $2.4$--$3.0\times$ higher per-joule efficiency on identical models (e.g., MiniMax-M2.5 on TBv2: $17.30$ vs.\ $5.70 \times 10^{-5}$), while M4 Max achieves $3.7$--$3.8\times$ higher per-watt efficiency due to its $3.5$--$5.3\times$ lower power envelope, with only $\sim$2.2pp accuracy degradation across both benchmarks. The 12--15$\times$ higher M4 Max latency reflects the same memory-bandwidth bottleneck observed in single-turn evaluation, with multi-turn workloads amplifying absolute differences but preserving relative rankings.}
\label{tab:multiturn_efficiency}
\end{table}

The qualitative patterns from single-turn evaluation persist (Table~\ref{tab:multiturn_efficiency}): on identical models, cloud hardware achieves $2.4$--$3.0\times$ higher per-joule efficiency than local (e.g., MiniMax-M2.5 on TBv2: $17.30$ vs.\ $5.70 \times 10^{-5}$ IPJ), consistent with Tables~\ref{tab:apple_m4_max_vs_nvidia_b200_for_power}--\ref{tab:apple_m4_max_vs_nvidia_b200_for_energy}, while local hardware achieves $3.7$--$3.8\times$ higher per-watt efficiency due to its $3.5$--$5.3\times$ lower power envelope. Per-query accuracy on the strongest local model is within ${\sim}2.2$~pp of the cloud configuration on both benchmarks (MiniMax-M2.5 on TBv2: $37.5$ vs.\ $39.7\%$; on GAIA: $14.2$ vs.\ $16.4\%$), suggesting that the routing-based savings reported in Section~\ref{sec:efficiency_gains_from_routing} carry over qualitatively to agentic workloads. Absolute savings rates may shift because multi-turn workloads have substantially longer effective context lengths and more tool-call overhead than single-turn chat (visible here as per-query energies in the $kJ$ rather than $J$ range). We caution that $245$ multi-turn queries is a small evaluation and view this experiment as a sanity check rather than a definitive characterization; full multi-turn IPW characterization across more diverse agent stacks is an important direction for future work.

\end{document}